\documentclass[letterpaper]{article}
\usepackage{aaai2026}

\usepackage{times}
\usepackage{helvet}
\usepackage{courier}
\usepackage{placeins}
\usepackage{enumitem}
\usepackage[hyphens]{url}
\usepackage{graphicx}
\usepackage{natbib}
\usepackage{caption}
\usepackage{tabularx}
\usepackage{booktabs} 
\usepackage{amsmath}
\usepackage[utf8]{inputenc}
\usepackage{textcomp}
\usepackage{needspace}

\AtBeginDocument{%
  }

\begin{document}

\newcommand{\reactsucc}{\ensuremath{C_{reactive\_errorless}}}
\newcommand{\reactfail}{\ensuremath{C_{reactive\_errors}}}
\newcommand{\advonly}{\ensuremath{C_{adv\_only\_errorless}}}
\newcommand{\procsucc}{\ensuremath{C_{proactive\_errorless}}}
\newcommand{\procfail}{\ensuremath{C_{proactive\_errors}}}
\newcommand{\advonlyerr}{\ensuremath{C_{adv\_only\_errors}}}

\newcommand{\showcomments}{1}
\newcommand{\roya}[1]{\ifnum\showcomments=1 \textcolor{brown}{[Roya: #1]}\fi}

\title{Asymmetric Trust Effects of Corrective AI in Expert Advisory Workflows under Epistemic Dependence}

\author {
    Dennis Kim,
    Roya Daneshi,
    Bruce Draper,
    Sarath Sreedharan
}
\affiliations {
    Colorado State University\\
    d.kim@colostate.edu,
    roya.daneshi@colostate.edu, 
    Bruce.Draper@colostate.edu,
    sarath.sreedharan@colostate.edu
}

\maketitle

\begin{abstract}
The increasing integration of AI-powered tools into expert workflows, such as medicine, law, and finance, raises a critical question: how does AI involvement influence a user’s trust in the human expert, the AI system, and the human-AI team? This question is especially important in expert advisory settings where users are epistemically dependent on human-AI systems: they are recipients of guidance produced by an expert using AI support, but often lack the domain knowledge needed to independently verify the recommendation. We investigated these dynamics through a user study ($N=157$) using a simulated course-planning task. Our design varied advisor performance and the structure of AI involvement, including whether AI support was present and, when present, whether it was invoked by the advisor or automatically monitored the interaction. Across all conditions, workflows ultimately produced correct schedules. Results show an asymmetric trust effect: advisor errors significantly reduce trust in the human advisor, but visible AI correction does not produce a corresponding increase in trust toward the AI assistant. Trust judgments remain anchored to the advisor across multiple trust measures, and changing the visible structure of AI involvement does not substantially redistribute trust toward the AI assistant or the human-AI team. These findings suggest a limit of corrective AI as a governance mechanism: making AI oversight visible may improve workflow recoverability, but it does not necessarily redistribute trust or responsibility away from the human-facing expert. In epistemically dependent settings, correctness alone may be insufficient for trustworthy AI integration, as users may continue to assign responsibility to the human expert even when AI assistance visibly shapes the final outcome.
\end{abstract}

\section{Introduction}

Across domains such as healthcare, financial advising, legal consultation, and others, people depend on human experts to help them navigate complex decisions~\cite{agrawal2023crosscultural, HanKo2025TrustDynamicsFA, mayer1995integrative, kaye2004roles}. These interactions depend on trust in experts~\cite{hendriks2015meti}: an evaluative judgment that an advisor will help achieve a user’s goals under conditions of uncertainty~\cite{lee2004trust}, which in practice often manifests in reliance behaviors such as reuse~\cite{liao2022designing}. Understanding how trust is shaped within human-AI workflows is therefore critical for interpreting user perceptions and anticipating downstream reliance behavior.

As AI systems increasingly enter expert advisory workflows, users may attribute guidance to the human advisor, the AI system, or the human-AI workflow as a whole. This creates a puzzle for corrective AI: when AI helps identify or resolve an expert error, does it earn trust as a reliable contributor, or does its intervention primarily expose the human expert’s fallibility? Even before a specific intervention occurs, the presence of AI may also shape how users interpret credibility within the advisory process~\cite{nakano2025understanding}. The implications of this shift in advisory contexts remain unclear, and understanding how responsibility and epistemic authority are allocated within such human-AI advisory workflows is central to broader questions of accountability and governance in sociotechnical systems.  The “user” refers to the recipient of guidance (e.g., a patient), who does not directly interact with the AI but evaluates the outcomes of a human-AI workflow. As such, the user serves as the trustor, forming judgments about the advisor, the AI system, and their combined behavior. 

\begin{figure*}[tbp]
  \centering
  \includegraphics[width=0.32\linewidth]{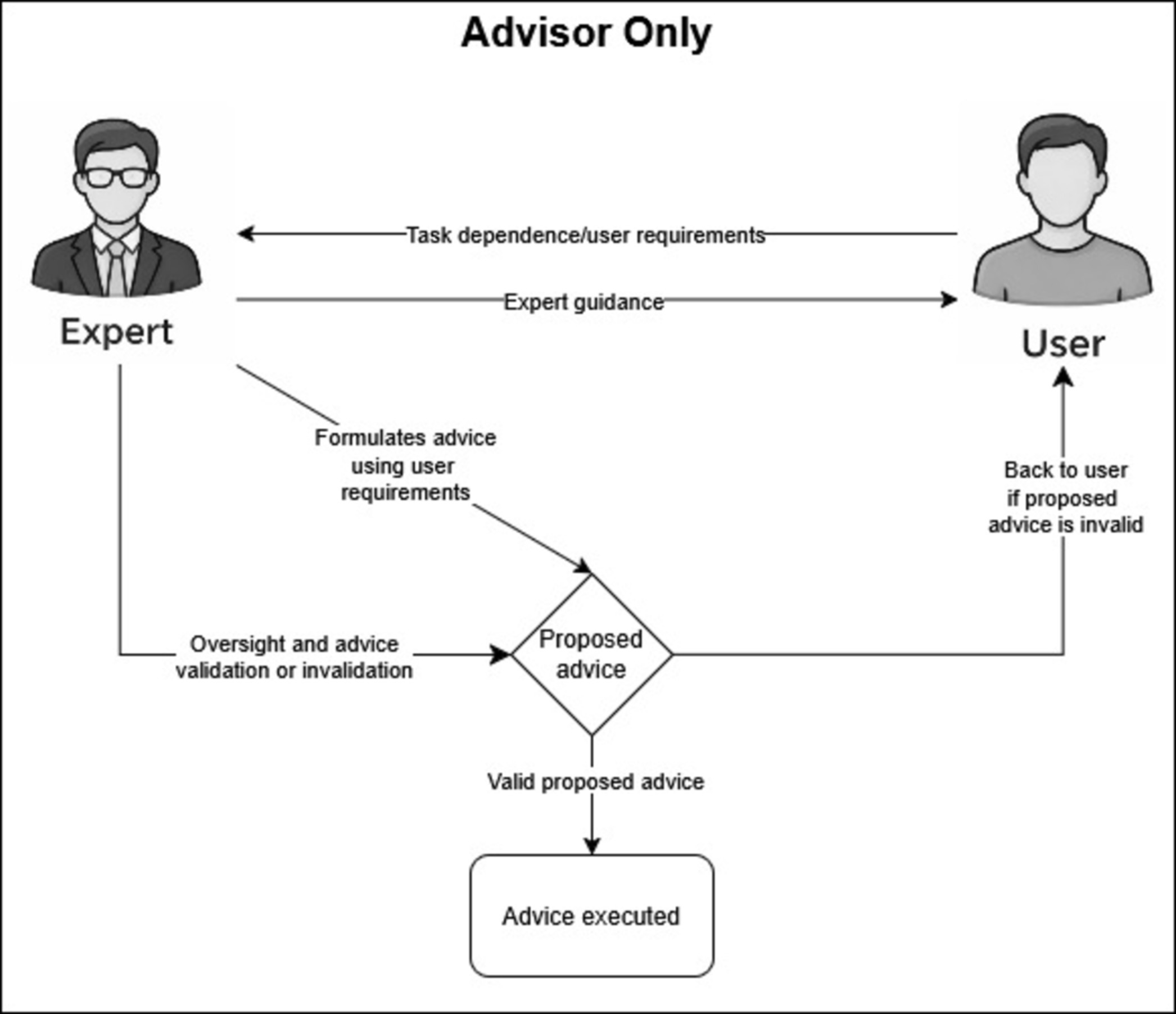}\hfill
  \includegraphics[width=0.32\linewidth]{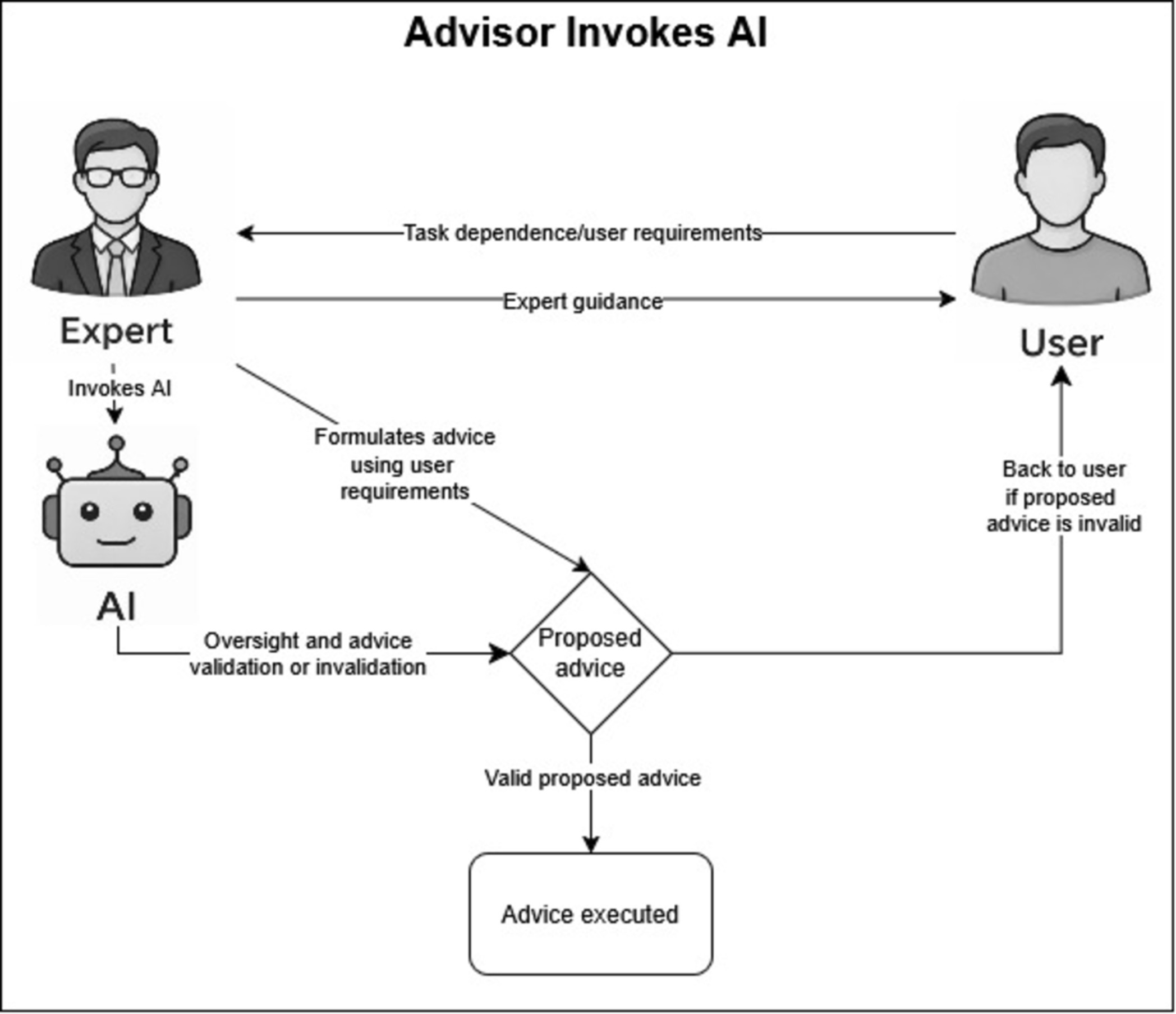}\hfill
  \includegraphics[width=0.32\linewidth]{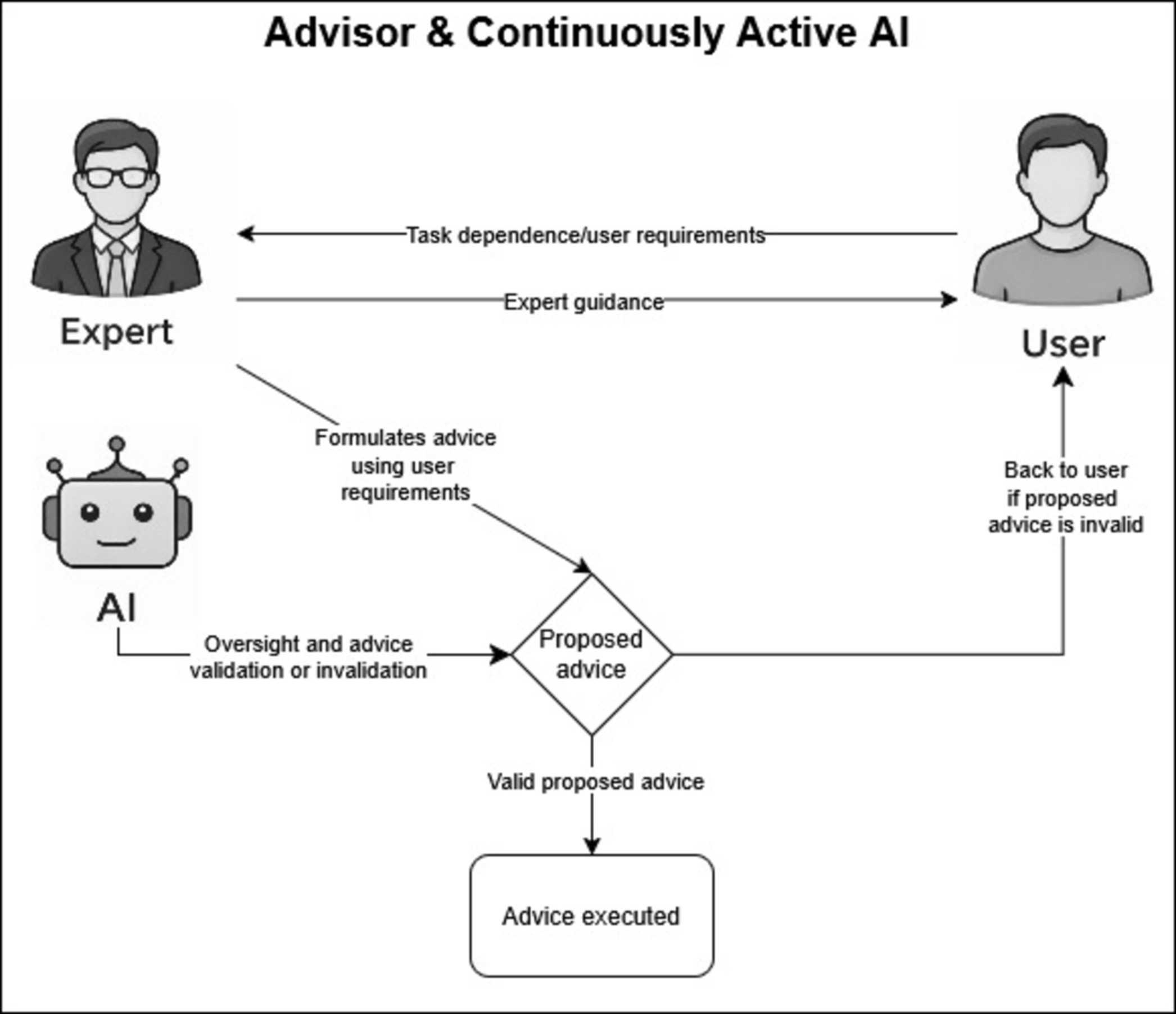}
  \parbox[t]{0.32\linewidth}{(a)\label{fig:adv_ai_auto_a} Advisor-only workflow without AI involvement ($C_{adv\_only\_errorless}$ or $C_{adv\_only\_errors}$).}\hfill
  \parbox[t]{0.32\linewidth}{(b)\label{fig:adv_ai_invoked_a} Advisor-initiated AI correction following an error ($C_{proactive\_errorless}$ or $C_{proactive\_errors}$).}\hfill
  \parbox[t]{0.32\linewidth}{(c)\label{fig:adv_only_a} Automatic AI oversight and correction ($C_{reactive\_errorless}$ or $C_{reactive\_errors}$).}
  
  \caption{Illustrative advising workflows studied, highlighting differences in advisor performance and AI intervention modality.}
  \label{fig:intro_workflows}
\end{figure*}

Prior work has examined trust in algorithmic systems~\cite{lee2004trust,lee2018understanding} and human experts~\cite{hendriks2015meti}, but less is known about how users distribute trust across human and AI contributors in precision-dependent tasks where both shape the final guidance. This gap is especially important when AI visibly corrects or safeguards expert recommendations, because such interventions may change how users assign credibility and responsibility within the advisory workflow.

From an epistemic perspective, trust in experts can be decomposed into expertise, integrity, and benevolence~\cite{hendriks2015meti}, with advisor performance informing perceived expertise~\cite{tenney2008calibration} and AI oversight potentially shaping integrity- and benevolence-related inferences. This multidimensional approach remains well-justified in trust measurement~\cite{kim2024XAItrust} and is especially relevant for understanding trust in expert AI contexts~\cite{asan2020aiTrustHealthcare}, where AI integration is accelerating~\cite{kaffee2025local,moell2025medical}. In this work, we experimentally manipulate two factors: advisor performance (whether the advisor fails to detect an invalid course selection) and AI oversight modality (advisor-only with no AI involvement, reactive AI oversight that automatically monitors and flags issues, or proactive AI oversight invoked at the human advisor’s discretion). We examine whether and how these factors influence multidimensional trust judgments, entity-specific trust, and downstream behavioral intentions in expert-guided interactions. As we show, these effects are not uniformly distributed across system components, but instead are concentrated in trust judgments directed at the human advisor.

We leveraged an online academic advising simulation as a simplified, representative task to study this phenomenon. Throughout this paper, we refer to the human expert as the advisor (or simply the expert), following the academic advising scenario used in our study. Figure~\ref{fig:intro_workflows} shows the interaction structures considered in our study. Academic advising offers a useful environment for studying these dynamics, since advisors are expected to have expertise and specialized knowledge about courses as well as departmental and university degree requirements~\cite{edutechloft2025_student_advising}. At the same time, while perceived risk is a necessary condition of trust~\cite{mayer1995integrative}, the stakes of advising are moderate enough to limit confounding factors present in higher-risk domains such as medicine or finance, enabling a more controlled examination of trust dynamics. We therefore frame this study, in part, as a mechanism-focused investigation of trust attribution under AI oversight rather than as a direct claim about trust dynamics in all expert domains. Moreover, the scenario naturally supports AI assistants that can run in both a proactive and a reactive mode (Section~\ref{subsec:study-design} provides a full overview of the experiment setup). Each measurement draws on previously validated trust instruments, as described in Section~\ref{subsec:measures}.

Our results show that advisor performance significantly
shapes trust-related evaluations: error-present conditions
were evaluated less favorably across measures of epistemic
trustworthiness, entity-specific trust, general trust, and
likelihood of advisor reuse, even though errors were
ultimately corrected in all conditions, either by the AI system
or the advisor. In contrast, AI intervention modality did not
produce significant effects, suggesting that visible AI
involvement did not substantially redistribute trust away
from the human advisor. Descriptively, performance-related
declines were largest in advisor-only conditions, followed by
proactive AI use, and smallest under reactive AI oversight.
Trust in the AI system remained comparatively stable, with
mixed results for the human-AI team.

Taken together, our results suggest that corrective AI may change how an advisory workflow reaches a valid outcome without changing who users treat as epistemically responsible for that outcome.

\subsection{Contributions}
\label{subsec:contributions}

To summarize, the paper makes three contributions:
\begin{enumerate}
    \item We present an interactive experimental paradigm for studying trust attribution in AI-assisted advisory workflows under controlled variations in advisor performance and AI integration structure.

    \item Using this academic advising setup, we show that observed advisor errors reduced trust in the human advisor, while visible AI correction did not increase trust in the AI assistant despite its corrective role. AI oversight also did not significantly redistribute trust consequences across the advisor, AI assistant, and advisor-AI team, suggesting that epistemic responsibility remained anchored to the human-facing expert. Descriptive patterns suggest that AI involvement may attenuate trust penalties, with heterogeneous responses to advisor mistakes.

    \item We derive design implications for human-AI advisory workflows, emphasizing the importance of advisor competence, real-world constraints on error visibility, the positioning of AI assistance in error detection, and the potential for complementary signals to support adaptive responses to user trust.
\end{enumerate}

\section{Hypotheses}
\label{subsec:hypotheses}

We test four hypotheses motivated by prior work on trust in automation~\cite{lee2004trust, Dzindolet2003Trust, hoff2015trust} and our directional expectations about how expert missteps and recovery influence trust. Conditions are defined by AI modality (none, proactive, reactive) and advisor performance (errorless, error-present), yielding six conditions: \advonly{}, \advonlyerr{}, \procsucc{}, \procfail{}, \reactsucc{}, and \reactfail{} (see Section~\ref{subsec:study-design}).

In all conditions, the student (participant) is given a course scheduling task (i.e., preparing a course schedule for a new semester), specifies desired courses, and receives a valid schedule. In the no-AI condition, the advisor completes the task unaided (Figure~\ref{fig:intro_workflows}(a)). In proactive conditions, the advisor invokes the AI to verify the scheduling choice, and errors, if present, are identified during verification (Figure~\ref{fig:intro_workflows}(b)). In reactive conditions, the AI automatically intervenes and immediately flags errors (Figure~\ref{fig:intro_workflows}(c)). When no error occurs, the interaction proceeds without an error-correction event. An advisor error refers to the advisor's initial failure to detect an invalid course selection, all such errors were corrected before the final schedule was produced. 

Dependent variables captured four trust-related outcomes: epistemic trustworthiness assessed via the \citet{hendriks2015meti} questionnaire; entity-level trust in the advisor, AI assistant, and human-AI team adapted from \citet{riedl2024patients}; global trust in the advisor on a 100-point scale; and willingness to reuse the advisor on a 7-point Likert scale (``Extremely unlikely''--``Extremely likely''). All hypotheses concern how advisor performance and AI intervention modality influence these evaluations.

Prior work on trust in automation suggests that observed failures prompt trust recalibration~\cite{Dzindolet2003Trust}. Accordingly, we expect advisor performance (error-present vs errorless) to influence trust-related evaluations of the advisor (H1). We further expect that AI intervention modality (none, proactive, reactive) will influence how trust-related evaluations are allocated across the advisor, AI assistant, and advisor-AI team (H2), consistent with frameworks emphasizing how agency and decision authority are distributed in human-automation workflows~\cite{Parasuraman2000TypesLevelsAutomation}. In this framework, differences in who initiates action (e.g., monitoring and verification) reflect different allocations of decision authority, which may influence how observers interpret an advisor’s epistemic responsibility even when task outcomes are identical. Because users can evaluate human, AI, and hybrid trust targets separately~\cite{riedl2024patients}, we expected visible AI verification to shift some trust credit from the advisor toward the AI assistant or advisor-AI team. This shift may be strongest under reactive oversight, where monitoring occurs independently rather than through advisor invocation.
We also expected AI modality to shape the trust penalty associated with advisor errors (H3), because communicated feedback and recovery can affect trust responses~\cite{Dzindolet2003Trust,hoff2015trust}. Whereas an unaided error primarily signals advisor fallibility, AI correction may signal that the workflow contains safeguards. Advisor-invoked verification may indicate responsible tool use, while automatic monitoring may indicate recovery independent of the advisor recognizing the need for assistance. We therefore expected AI oversight, particularly reactive oversight, to attenuate the error-related trust decline relative to advisor-only workflows. Finally, we expected advisor errors to reduce willingness to reuse the advisor (H4).

\subsection{Hypotheses List}
\label{hypotheses}

To summarize, the hypotheses we tested are as follows:
\begin{itemize}
    \item H1: Advisor errors will significantly negatively affect trust evaluations of the advisor, including epistemic trustworthiness measured using the \citet{hendriks2015meti} instrument and entity-specific trust adapted from \citet{riedl2024patients}.    
    
    \item H2: AI intervention modality (none, proactive, reactive) will significantly affect how trust-related evaluations are distributed across available trust targets, with AI-supported workflows expected to produce higher evaluations of the AI assistant and advisor-AI team when those targets are present.

    \item H3: The trust penalty from advisor errors will differ significantly depending on the AI intervention modality.

    \item H4: Participants will be significantly less likely to reuse advisors who made errors than those who did not, consistent with prior work linking trust to downstream reliance.
\end{itemize}

As reported below and in Section~\ref{sec:results}, H1 and H4 were supported: advisor performance significantly influenced advisor-directed trust evaluations across multiple measures, including epistemic trustworthiness~\cite{hendriks2015meti}, entity-specific trust~\cite{riedl2024patients}, general trust~\cite{lee2004trust}, and willingness to reuse~\cite{liao2022designing}. H2 and H3 were not supported: AI intervention modality did not produce significant main effects, nor did it significantly moderate the trust penalty associated with advisor errors.

We treat these deviations from the a priori expectations as theoretically informative, suggesting that observed advisor error played a more dominant role than AI involvement strategy in shaping trust and reuse judgments.

\section{Relevant Work}

Research on trust in automation, human-AI interaction, and social epistemology
provides complementary perspectives motivating our study design and trust
measurement strategy. We organize prior work around three questions central to our investigation:
(i) how trust responds dynamically to observed performance and error,
(ii) how procedural structure and responsibility cues shape trust in
AI-supported expert workflows, and (iii) how trust should be measured when
multiple potential trustees are present in a single interaction.

\subsection{Trust Calibration Under Epistemic Dependence}

A central theme in the trust-in-automation literature is trust recalibration: users update their trust judgments based on observed system behavior and outcomes \cite{lee2004trust}. Here, trust recalibration refers to differences in reported trust between error-present and errorless conditions following a single between-subjects interaction. This process is especially important under epistemic dependence, where users cannot fully understand or independently verify system correctness and must instead rely on observable cues such as performance, communication, and interaction dynamics to calibrate trust~\cite{lee2004trust}. Recent work in human-AI interaction further formalizes trust in terms of user vulnerability and expectations over system behavior~\cite{jacovi2021formalizing}, while work on trust dynamics emphasizes that trust judgments emerge over interaction rather than from isolated assessments~\cite{HanKo2025TrustDynamicsFA}.

Empirical evidence reinforces that calibration operates at the level of individual interactions and outputs. For example, \citet{kim2023humans} show that users do not rely on AI systems with a fixed level of trust, but instead selectively accept or reject outputs after engaging in verification behaviors, highlighting the dynamic and context-sensitive nature of trust. Prior syntheses similarly emphasize that trust judgments depend on contextual factors such as task demands, perceived risk, and user expectations, and may diverge from objective performance metrics~\cite{hoff2015trust,Wong2025TrustAutomationAIHealthcare}. Work on AI-assisted ethical decision-making further shows that reliance on AI advice can vary with properties of the recommendation itself, including whether it aligns with the decision-maker's values~\cite{narayanan2023value}. Across domains, observed errors play a particularly important role in this process, as they provide salient evidence that can prompt users to update their beliefs about system reliability and competence. For example, work on algorithmic aversion shows that exposure to system errors can reduce subsequent reliance after users observe imperfections~\cite{dietvorst2015algorithm}.

Importantly, trust responses depend not only on error occurrence but also on how errors are communicated and contextualized. \citet{Dzindolet2003Trust} show that feedback about system limitations and error characteristics enables more appropriate calibration, helping users avoid both overreliance and disuse. Work on AI communication similarly shows that factuality-expression formats can affect user trust in AI-generated answers~\cite{do2025hide}, while uncertainty-expression cues can modulate reliance by signaling limitations in system outputs~\cite{kim2024uncertainty}. Unlike settings where an AI system directly communicates reliability or uncertainty information to users, our study examines AI assistance embedded within a human-facing expert workflow. In this setting, corrective AI may provide evidence of workflow-level recovery while also exposing the advisor’s need for correction, without necessarily making the AI assistant itself the primary target of trust. This motivates examining outcomes alongside how performance information is surfaced, interpreted, and attributed within an interaction.

Recent studies in AI-assisted advisory contexts further characterize trust as path-dependent and sensitive to sequences of interaction events. \citet{HanKo2025TrustDynamicsFA} show that salient errors can produce sharp declines in trust that may be partially repaired through subsequent corrective interactions, while \citet{Pareek2024TrustDevelopmentRepair} demonstrate that trust can propagate across stages of a workflow when users are able to evaluate some components but not others. These findings highlight that trust judgments are formed over trajectories of interaction, where both errors and corrective signals contribute to how responsibility and reliability are inferred over time.

Together, these studies position trust calibration as an interaction-level process shaped by observed outcomes, communicated performance information, corrective signals, and the user’s ability to evaluate what happened. This is especially important in epistemically dependent advisory settings. Building on this literature, our work examines whether trust judgments are shaped not only by final outcomes, but by whether the advisory workflow visibly encountered and resolved problems along the way, and by how AI support is incorporated into that sequence.

\begin{figure*}[tbp]
  \centering
  \includegraphics[width=0.32\linewidth]{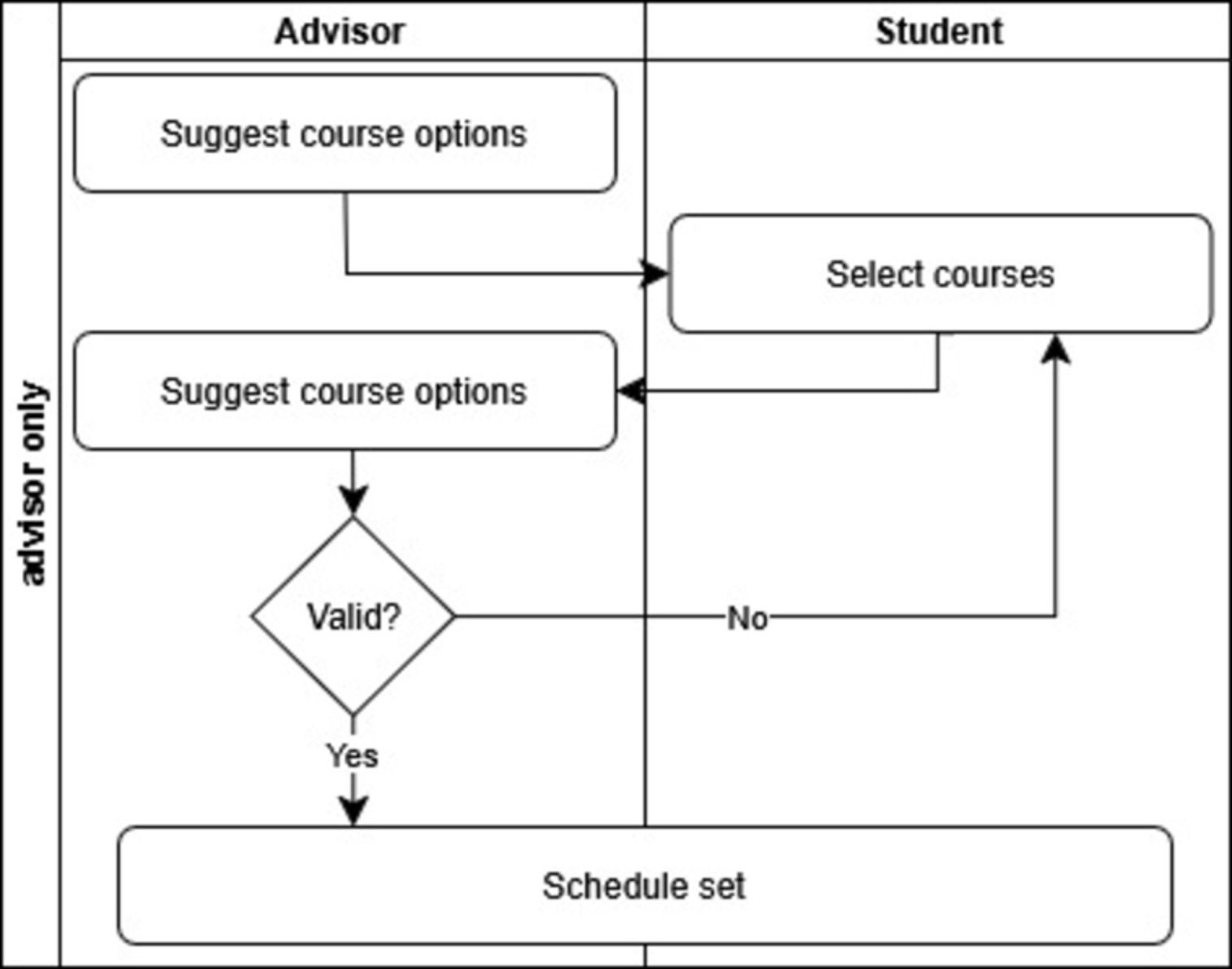}\hfill
  \includegraphics[width=0.32\linewidth]{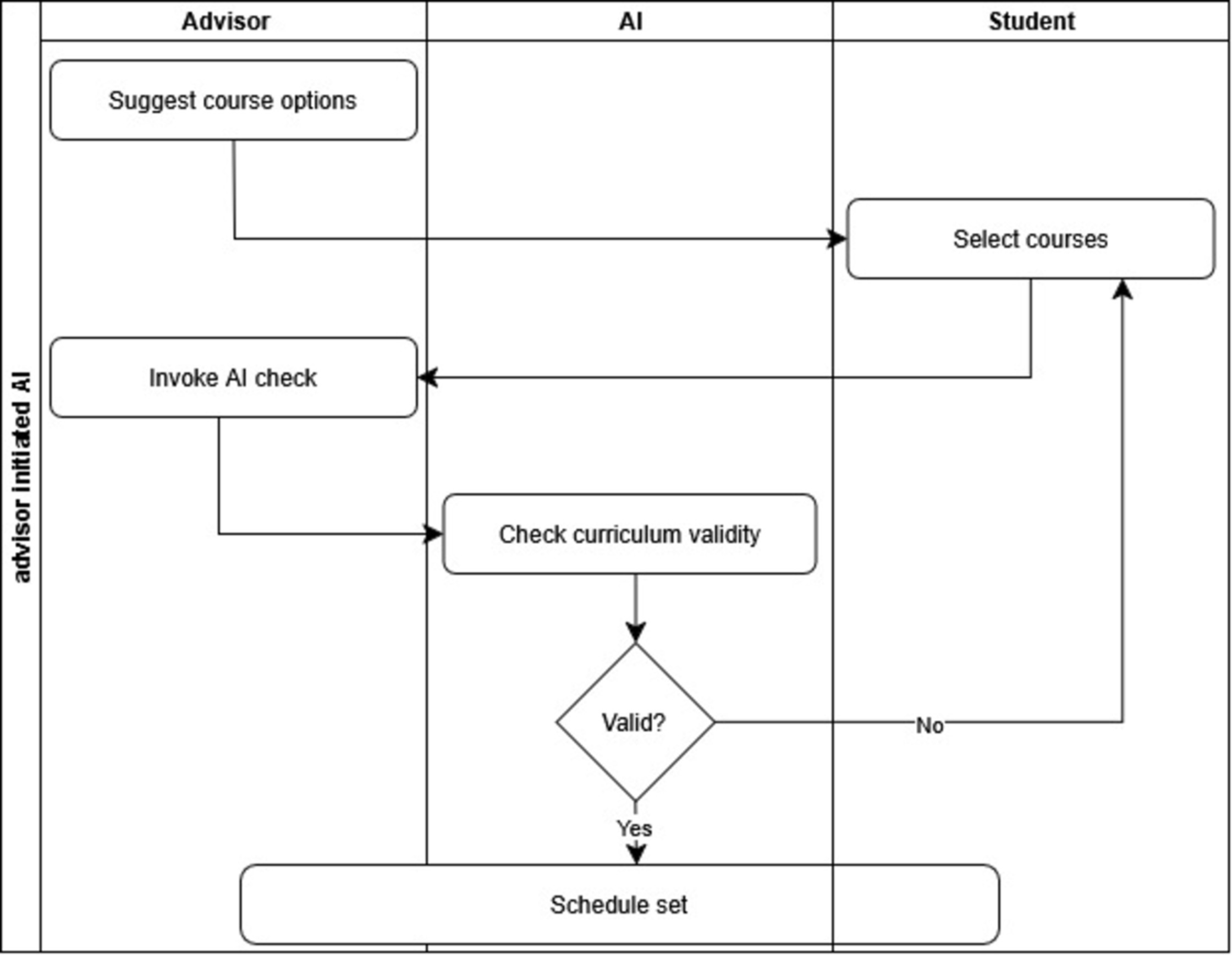}\hfill
  \includegraphics[width=0.32\linewidth]{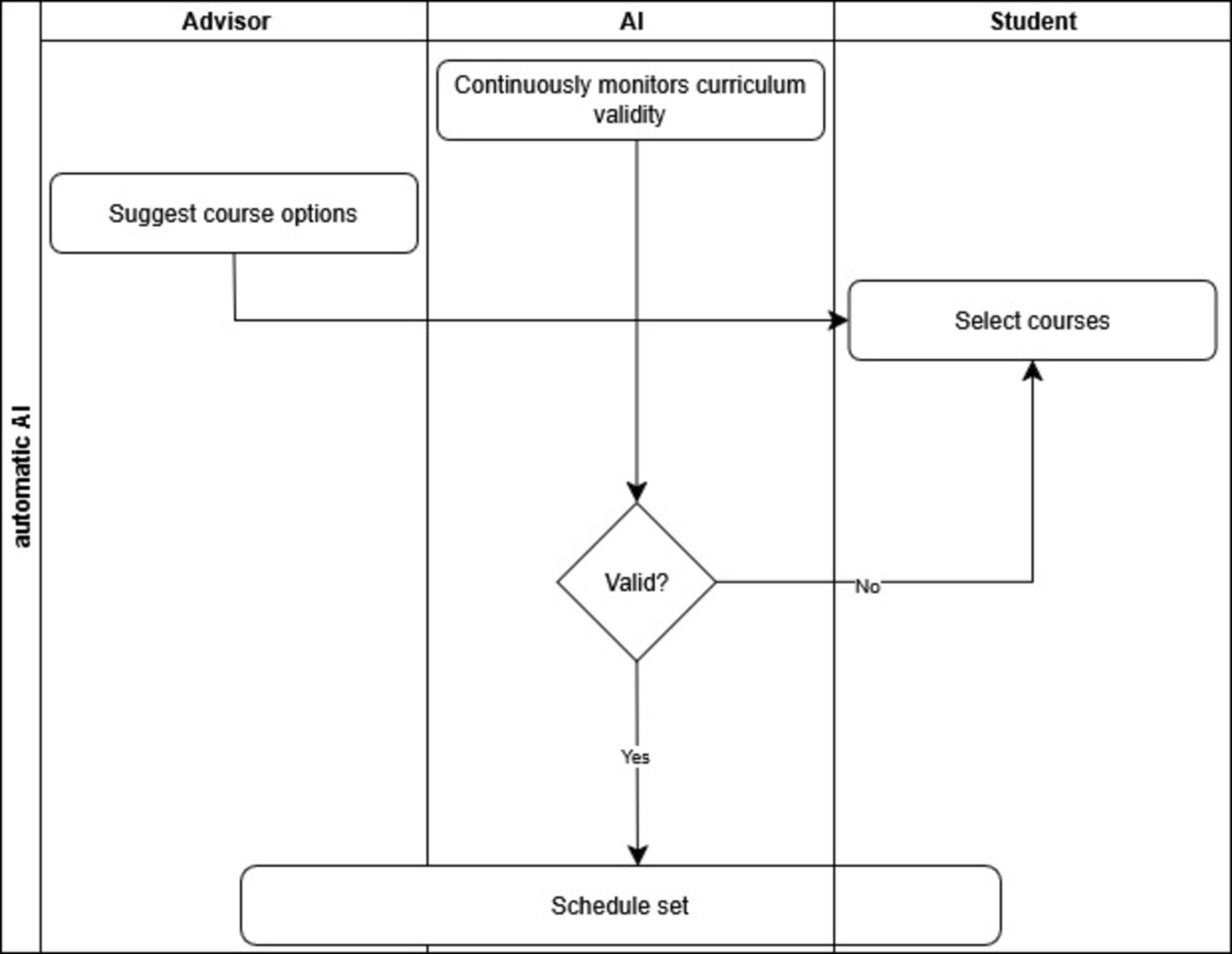}
  
  \parbox[t]{0.32\linewidth}{(a)\label{fig:adv_only} $C_{adv\_only\_errorless}$ or $C_{adv\_only\_errors}$: advisor-only workflow without AI involvement.}\hfill
  \parbox[t]{0.32\linewidth}{(b)\label{fig:adv_ai_invoked} $C_{proactive\_errorless}$ or $C_{proactive\_errors}$: advisor-initiated AI assistance following a detected error.}\hfill
  \parbox[t]{0.32\linewidth}{(c)\label{fig:adv_ai_auto} $C_{reactive\_errorless}$ or $C_{reactive\_errors}$: automatic AI oversight that continuously monitors and intervenes.}
  
  \caption{Interaction workflows for the academic advising paradigms examined in this study.}
  \label{fig:workflows}
\end{figure*}

\subsection{AI Oversight, Workflow Governance, and Responsibility Attribution}

Beyond performance outcomes, trust judgments are also shaped by how users
interpret epistemic roles and responsibility within a workflow. In human-AI
settings, identical outcomes may support different inferences depending on
who appears to monitor, verify, and authorize decisions. \citet{YangMa2025EpistemicRelationships} argue that users’ evaluations of AI systems and AI-assisted decision-making contexts involve context-dependent assessments of epistemic authority, responsibility, system behavior, and underlying processes. A parallel distinction appears in formal work on AI trustworthiness: \citet{bisconti2024formal} distinguish the characteristics a system possesses from the characteristics observers infer from displayed behavior, and argue that perceived trustworthiness can depend on cues such as transparency, agency locus, and human oversight. In our setting, these process-based assessments are reflected through procedural cues in the
interaction (e.g., how AI intervention occurs).

Classic models of human-automation interaction similarly emphasize the
importance of procedural visibility. \citet{Parasuraman2000TypesLevelsAutomation}
note that higher levels of automation can obscure ongoing human involvement,
reducing users’ ability to distinguish between human and system contributions.
Such shifts do not inherently undermine trust, but they may alter how users
attribute responsibility for errors and corrections.

How AI involvement is disclosed may also shape attribution. Progressive disclosure reveals explanations gradually and on demand, with their order, modality, and complexity affecting perceived transparency and understanding~\cite{muralidhar2025progressive}. Thus, trust attribution may depend not only on whether AI correction is visible, but also on when and how its process is explained.

Social epistemology suggests that trust is supported by
mechanisms that evaluate an informant’s competence and responsiveness to error
rather than by passive acceptance alone~\cite{Sperber2010EpistemicVigilance,Origgi2014EpistemicTrust}.
Observable signals of monitoring and correction can therefore function as
procedural evidence of epistemic responsibility. Related work on calibration
and credibility judgments shows that errors are particularly damaging when interpreted as evidence of poor self-assessment rather than as
isolated outcomes~\cite{kim2004removing,tenney2008calibration,GriffinTversky1992}. Together, these perspectives motivate treating AI oversight as part of advisory workflow governance rather than as a purely technical aid. Building on this, we examine whether visible AI correction in a human-facing advisory workflow changes how trust consequences are allocated across the advisor, the AI assistant, and their combined operation, even when the final recommendation is correct.

\subsection{Trust, Reliance, and Accountability Across Multiple Targets}

A methodological challenge in AI-supported advisory systems is that a single
interaction may involve multiple plausible trustees. Users may evaluate the
human advisor, the AI system, and their combined operation differently, and
outcome evidence may not map cleanly onto a single target.

\citet{hendriks2015meti} conceptualize epistemic trustworthiness
as a multidimensional construct comprising expertise, integrity, and
benevolence, providing a validated framework for assessing trust in experts
under epistemic uncertainty. However, this alone misses how trust distributes across multiple entities when AI is involved. Prior work~\cite{riedl2024patients}
demonstrates that users can meaningfully provide entity-specific trust
judgments for human professionals, AI systems, and hybrid configurations
using parallel item structures.

This multi-target measurement strategy is central to our study. By separately measuring trust in the advisor, the AI assistant, and the advisor-AI team, we can test whether trust penalties and credit are allocated across specific components of the workflow or remain concentrated on the human expert despite visible AI involvement. We use trust consequences to mean condition-related differences in trust and the entities toward which they are directed. Advisor-directed consequences were assessed through epistemic trustworthiness, advisor-specific trust, and general trust; AI- and team-directed consequences through entity-specific ratings in AI-supported conditions. Advisor reuse was analyzed separately as a behavioral proxy for anticipated reliance.

\section{Methodology}
\subsection{Study Design}
\label{subsec:study-design}

We examine how trust judgments form in response to realistic advising sequences and AI involvement. Each participant received a distinct, randomly-assigned workflow varying along two factors: advisor performance and AI intervention modality, to assess their effects and interaction.

Participants engaged in a simulated academic advising interaction (Fig.~\ref{fig:workflows}) presented as a sequence of screenshots and system messages within a web-based interface~\cite{qualtrics} (Appendix~\ref{app:sim-walkthrough} Figs.~\ref{fig:appendix-sim-seq}, ~\ref{fig:appendix-sim-seq-part2}). They were informed that they would complete a course scheduling task (with or without AI) and that the study examined perceptions of this interaction. The AI, when present, operated in an oversight tool. When errors occurred, they were observable: with AI, they were surfaced through messages and visual text feedback indicating invalid course selections; without AI, the advisor identified them by cross-referencing a rule book and addressed them in the chat. In all conditions, participants assumed the role of the student and were instructed to follow the guidance. They were never provided means to verify their choices independently. 

A between-subjects design was used to avoid carryover and learning effects, as trust judgments recalibrate based on prior performance~\cite{lee2004trust}. Each participant completed the task once. Post-task, participants evaluated the advisor, while those in AI-supported conditions also evaluated the AI assistant and advisor-AI team.


An a priori G*Power analysis of the advisor-performance $\times$ AI-modality interaction showed that the 3 $\times$ 2 design required $N=158$ to detect $f=.25$ with $\alpha=.05$ and 80\% power; the final $N=157$ provided approximately 79.9\% power. Under the same assumptions, the restricted 2 $\times$ 2 analyses required $N=128$; the available AI-supported subsample of $N=109$ provided approximately 73.4\% power.

The experimental interactions were as follows (illustrative workflows available in Fig.~\ref{fig:workflows}):

\begin{description}
  \item[\advonly{}:] Advisor-only, errorless condition. The student selects courses that satisfy the scheduling requirements, and the advising workflow proceeds without AI involvement.

  \item[\advonlyerr{}:] Advisor-only, error-present condition. The student selects an invalid course, which the advisor detects only after checking the rule book and corrects without AI involvement.

  \item[\procsucc{}:] Advisor-invoked AI, errorless condition. The student selects courses that satisfy the scheduling requirements within a workflow where AI support is available through advisor invocation.

  \item[\procfail{}:] Advisor-invoked AI, error-present condition. The student selects an invalid course that the advisor does not initially detect; advisor-invoked AI assistance identifies the error so it can be corrected before the final schedule is produced.

  \item[\reactsucc{}:] Automatic AI oversight, errorless condition. The student selects courses that satisfy the scheduling requirements within a workflow involving automatic AI oversight.

  \item[\reactfail{}:] Automatic AI oversight, error-present condition. The student selects an invalid course that the advisor does not initially detect; automatic AI oversight flags the error so it can be corrected before the final schedule is produced.
\end{description}

\subsection{Participants}

Participants were recruited through Prolific~\cite{prolific} and completed the study remotely. Repeat participation was systematically restricted using Prolific’s built-in eligibility controls to maintain data integrity. Participation was restricted to desktop or laptop users to ensure consistent interaction with the advising interface. In addition, participants were excluded from analysis if they failed the attention-check questions embedded throughout the study, including within the interactive simulation and the final questionnaire. No other exclusion mechanisms were applied. 

A total of 160 participants completed the study. After accounting for those failing the attention-checks, the final analytic sample was $N = 157$. Each was compensated \$4.50 for their time upon completion of the study. All study procedures were approved by the institutional review board, and informed consent was obtained from all participants. Participants were assigned to experimental interactions via Prolific’s randomization, and demographic information is summarized in the Appendix~\ref{app:participant-breakdowns}. The study took approximately 17 minutes to complete, including the advising simulation and post-task questionnaire.

\subsection{Measures}
\label{subsec:measures}

We intentionally employed multiple trust measures to capture distinct but complementary aspects of trust in human-AI advisory interactions. The \citet{hendriks2015meti} instrument was used to assess epistemic trustworthiness in the advisor, while items adapted from \citet{riedl2024patients} captured broader entity-specific trust judgments toward the advisor, AI assistant, and advisor-AI team. Consistent with \citet{lee2004trust}, these items capture trust as an evaluative attitude toward system entities. Finally, the single-item trust measure provided a holistic assessment of trust that was not decomposed into specific dimensions, allowing us to examine whether broader trust judgments aligned with or diverged from measures of specific characteristics.

We adapted items from~\cite{hendriks2015meti,riedl2024patients} for the academic advising scenario through minor wording changes when necessary (e.g., replacing domain-specific referents with ``advisor,'' ``AI assistant,'' or ``advisor-AI team'') while preserving the original response formats and construct intent. Unless otherwise noted, all multi-item measures were assessed using a 7-point Likert scale ranging from 1 (``Strongly [disagree/negative construct]'') to 7 (``Strongly [agree/affirmative construct]''). Questionnaire items from~\citet{riedl2024patients} and~\citet{hendriks2015meti} used in the study are reported in Appendices~\ref{app:riedl-items} and~\ref{app:meti-items}, respectively.

For multi-item constructs, we aggregate items within each instrument’s intended structure and report results at the composite level when appropriate. This approach reduces item-level noise and aligns inference with the theoretical level at which the constructs are defined (e.g.,~\citet{hendriks2015meti} expertise, integrity, and benevolence), while preserving the original response scale as originally structured. In addition, aggregating items according to validated instrument structures \cite{hendriks2015meti} reduces the number of statistical tests performed, thereby limiting family-wise error rate (FWER) inflation and the risk of spurious significance.

\subsubsection{Epistemic Trustworthiness (METI)}

To assess participants’ epistemic trust in the advisor, we administered items from the Muenster Epistemic Trustworthiness Inventory (METI)~\cite{hendriks2015meti}. METI operationalizes epistemic trustworthiness along three dimensions: perceived expertise, integrity, and benevolence, reflecting perceived competence, adherence to professional standards (e.g., honesty and fairness), and goodwill toward others, respectively. Following the instrument’s dimensional structure, we computed dimension scores by averaging item responses within each dimension (higher scores indicate stronger perceived epistemic trustworthiness). Item counts and internal consistency for each composite are reported in Appendix~\ref{app:measures}. The full METI item set and dimension groupings used in this study are provided in Appendix~\ref{app:meti}.

\subsubsection{Trust in the Advisor, AI System, and Team}

In addition to epistemic trustworthiness, we measured entity trust using three items adapted from \citet{riedl2024patients}. Items were adapted to the advising scenario. All participants rated the advisor with those in AI-supported conditions also rating the AI assistant and advisor-AI team. Advisor-only participants did not rate absent entities. An example item is: ``I can trust the information presented by the advisor.''  
All items demonstrated acceptable internal consistency as measures of entity trust (see Appendix~\ref{app:measures}).

\subsubsection{General Trust}

To capture a global assessment of trust that was not tied to a specific Likert instrument, participants completed a single-item general trust rating: ``From 1--100, how trustworthy would you rate your advisor? (1 being the least trustworthy and 100 being the most)''. This measure served as a holistic outcome capturing overall trust in the advisor, unconstrained by specific item framing. A continuous scale allowed for finer-grained variation in global trust judgments than would be possible with a coarse Likert scale.

\subsubsection{Likelihood of Advisor Reuse}

To assess behavioral intention, participants reported their likelihood of reusing the advisor in the future. This single-item outcome was analyzed separately from the multi-item trust composites.

We treat reuse intention as a behavioral proxy rather than a direct trust measure. Whereas trust instruments capture evaluative judgments (e.g., perceived expertise or trustworthiness), reuse intention reflects whether participants would act on those judgments by relying on the advisor again. Reuse intention is particularly relevant in advisory settings, where the practical consequence of trust is often continued reliance over time. This measure therefore assesses whether differences in epistemic or entity-level trust correspond to differences in anticipated reliance on the advisor.

\subsection{Limitations}
We took care in designing this simulation to approximate a realistic, though necessarily simplified, interaction involving AI-supported decision-making. Nevertheless, several limitations should be acknowledged. First, the use of a between-subjects design prevents individual participants from directly comparing different advising interactions or outcomes, which may limit insight into how trust recalibrates across repeated interactions.

As with all controlled experimental studies, the findings may also be limited in their generalizability beyond the simplified advising task, interface, and population studied. While we sought to enhance situational validity through a realistic advising scenario and grounded operationalizations of advisor performance and AI involvement, participants’ perceptions may differ in real-world advising contexts and especially in higher-stakes expert domains such as medicine, legal consultation, or finance. We therefore interpret the present results as evidence about trust attribution under conditions of epistemic dependence in a controlled proxy setting, rather than as a direct estimate of how strongly the same effects would appear in safety-critical environments.

A further limitation is that the AI was always correct when it intervened, allowing us to isolate advisor performance and intervention modality but not AI uncertainty or failure. This may have limited variation in AI-directed trust and contributed to the null AI-trust effects. Future work should vary AI reliability and examine how advisor and AI errors jointly shape trust in the advisor, AI assistant, and human-AI team.

Additionally, individual differences such as prior experience with academic advising, familiarity with AI systems, or general attitudes toward automation may have influenced trust judgments. These factors were not explicitly measured and therefore fall outside the scope of the present study. Future work could examine how such individual characteristics interact with advisor performance and AI involvement to shape trust, including whether they moderate which aspects of trust are affected and which remain stable.

Although our results suggest that responsibility attribution remained anchored to the human advisor, we infer this pattern from the distribution of trust judgments across the advisor, AI assistant, and advisor-AI team rather than from a direct measure of perceived responsibility or accountability. Future work should include explicit responsibility-attribution measures to test this interpretation more directly.

\section{Results}
\label{sec:results}

\subsection{Overview of Analytical Approach}

Advisor-directed outcomes were analyzed using 3 $\times$ 2 factorial ANOVAs crossing advisor performance (error vs.\ errorless) with AI modality (none, proactive, reactive). Because AI and advisor-AI-team ratings were collected only in AI-supported conditions, those outcomes used 2 $\times$ 2 factorial ANOVAs crossing performance with proactive versus reactive modality. These analyses evaluated the hypotheses: advisor performance effects tested H1 and H4, AI modality main effects tested H2, and advisor performance by modality interactions tested H3. Outcomes included epistemic trustworthiness composites (expertise, integrity, benevolence~\cite{hendriks2015meti}), entity-level trust adapted from~\citet{riedl2024patients}, a 1 to 100 general trust rating, and likelihood of advisor reuse. Tukey’s HSD pairwise comparisons were conducted as exploratory analyses across all condition combinations. Effect sizes are reported as $\eta^2$ and Cohen’s $d$, with test statistics in standard ANOVA form and exact degrees of freedom. Variance differences between error-present and errorless conditions were assessed using Levene’s test.

\subsection{Omnibus Tests Across Trust Constructs}
\label{subsec:omnibus}

We first report the full 3$\times$2 factorial analyses, which provide the primary basis for evaluating the hypotheses. Complete omnibus statistics are reported in Appendix~\ref{app:omnibus-tests}.

\textbf{Advisor-directed trust~\cite{riedl2024patients}.} Advisor performance significantly affected all advisor-directed trust measures, with error-present conditions receiving lower evaluations than errorless conditions for perceived trustworthiness, information trustworthiness, and affective comfort. General trust showed the same pattern. Likelihood of advisor reuse was also lower in error-present conditions. These results support H1 and H4.

\textbf{Epistemic trustworthiness~\cite{hendriks2015meti}.} Advisor performance significantly affected
perceived expertise, integrity, and benevolence, with higher
evaluations in errorless than error-present conditions. These
results further support H1.

\textbf{AI-supported conditions (2$\times$2 analysis)~\cite{riedl2024patients}.} Restricting analyses to AI-supported conditions showed the same performance-driven pattern for advisor-directed trust measures. Advisor-AI team evaluations showed weaker and less consistent effects, while AI assistant evaluations alone were not significantly affected by advisor performance.

No significant main effects of AI intervention modality were observed across outcomes, so H2 was not supported. No significant advisor performance by modality interactions were observed, so H3 was not supported. To further evaluate these null effects, we conducted Bayesian ANOVAs on AI-supported trust evaluations using JASP's default JZS Bayes factor approach with Cauchy priors. These analyses provided evidence favoring exclusion of AI modality effects ($BF_{\mathrm{excl}} = 3.66$--$6.74$) and modality $\times$ performance interactions ($BF_{\mathrm{excl}} = 3.62$--$14.53$) across AI assistant, advisor-AI team, and advisor-directed trust measures (Appendix~\ref{app:bayesian-anovas}).

\subsection{Descriptive Statistics}

Descriptive statistics for all trust constructs are reported in Appendix~\ref{app:descriptive_tables}. Error-present conditions were evaluated less favorably across advisor measures, consistent with H1 and H4. AI modality differences showed no consistent pattern, consistent with unsupported H2 and H3. Error-present conditions also exhibited greater variability for several measures (e.g., expertise). Significant Levene’s test results were observed only for advisor-directed measures, with no significant variance differences for the AI assistant or combined team (Appendix~\ref{app:variance_tables}). These exploratory variance analyses were not primary tests of the hypotheses.

\subsection{Pairwise Condition Comparisons}

Exploratory Tukey HSD comparisons, with factorial ANOVAs serving as the primary basis for inference, are reported with full statistics in Appendix~\ref{app:pairwise-comparisons}. Epistemic expertise was lower in \advonlyerr{} than in \advonly{}, and in \procfail{} than in \procsucc{} and \advonly{}. Affective comfort was lower in \advonlyerr{} than \advonly{}, \reactsucc{}, and \procsucc{}, while \procfail{} was lower than \reactsucc{} and, in the AI-supported analysis, \procsucc{}. Information trustworthiness was lower in \advonlyerr{} than in \reactsucc{}, and in \procfail{} than in \advonly{}, \reactsucc{}, and, in the AI-supported analysis, \procsucc{}. Advisor reuse and general trust were each lower in \advonlyerr{} than in \advonly{}, \reactsucc{}, and \procsucc{}, and lower in \procfail{} than in \advonly{} and \procsucc{}.

Overall, significant contrasts primarily reflected lower evaluations in error-present than in errorless conditions, supporting H1 and H4. They did not reveal a consistent AI-modality effect or advisor-performance-by-modality interaction supporting H2 or H3.

\subsection{Summary of Results}

Across analyses, advisor errors reduced advisor-directed trust evaluations, including epistemic trustworthiness, entity-specific trust, general trust, and advisor reuse, supporting H1 and H4. AI modality did not significantly change trust evaluations of the advisor, AI assistant, or advisor-AI team, and advisor performance effects did not differ across modalities. H2 and H3 were not supported by the frequentist analyses. Bayesian ANOVAs further provided evidence favoring exclusion of AI modality effects and modality $\times$ performance interactions across AI-supported trust evaluations (Appendix~\ref{app:bayesian-anovas}). The 3$\times$2 and restricted 2$\times$2 interaction analyses had approximately 79.9\% and 73.4\% power, respectively, to detect $f=.25$; smaller interactions cannot be ruled out. Pairwise comparisons primarily reflected error-present versus errorless contrasts.

\section{Discussion}
\label{sec:discussion}

This study examines how advisor performance and AI involvement shape trust judgments in an interactive academic advising task. The design allows us to distinguish performance-driven trust calibration from attributional effects associated with AI intervention modality.

\subsection{Trust Calibration to Performance and Modality}

Although all conditions produced a correct schedule, participants responded meaningfully depending on whether the advisor failed to detect an invalidating course choice. AI modality did not produce reliable factor-level effects, suggesting that perceptions of advisor performance were the most salient factor shaping trust evaluations. The observed pattern aligns with prior work on trust calibration~\citep{Dzindolet2003Trust}, but in this setting these effects were concentrated on the human advisor rather than the AI assistant. 

The reuse-intention results extend this pattern from evaluative trust judgments to anticipated reliance behavior. Although we do not directly model the relationship between entity-specific trust and reuse, the parallel decline in reuse intention suggests that performance-related trust penalties may have downstream implications for continued reliance.  

Descriptive mean patterns suggest that AI involvement may shape evaluations in limited ways when errors are present, as performance-related declines followed a consistent ordering across modalities and multiple dependent variables, though these effects were not statistically significant at the factor level. Comparing average evaluation drops between errorless and error-present conditions generally showed the largest penalties in advisor-only workflows, followed by advisor-invoked AI workflows, and the smallest penalties under automatic AI oversight (Appendix~\ref{app:descriptive_tables}). These descriptive patterns suggest a possible avenue for future work: modality differences may become more salient in higher-stakes contexts, where advisor-invoked AI could signal uncertainty differently than automatic AI oversight.

Beyond mean differences, failure conditions were associated with greater variability in trust responses, with standard deviations consistently higher in error conditions than in their errorless counterparts, particularly for advisor-directed measures where Levene’s tests indicated significant differences. This pattern is detailed in Appendix~\ref{app:variance_tables}. This may reflect differences in how participants tolerate failure, the extent to which these errors violated expectations, or potential compounding effects from repeated errors. However, identifying the underlying driver of this pattern was not part of our experimental design and is left for future work. 

\subsection{Trust Attribution and the Effects of AI Involvement}

A central pattern in our results is that corrective AI changed the structure of the advisory workflow without significantly changing the target of trust consequences. Observed errors produced an attributional form of trust calibration, with trust consequences remaining anchored to the human advisor rather than being redistributed across the human-AI advisory system. Even when the AI assistant visibly detected and corrected errors, participants did not assign corresponding trust credit to the AI assistant. One interpretation is that AI correction creates a double signal: it demonstrates that the advisory workflow can recover from errors, but it also reveals that the human-facing advisor required correction. Under epistemic dependence, users may therefore treat the advisor as the primary bearer of responsibility, rather than crediting the AI assistant that contributes to the correction. In our setting, AI intervention modality defines how errors are detected and corrected, effectively acting as a governance mechanism over the advisory workflow. Across these modalities, verification authority is distributed differently. In advisor-only workflows, authority appears concentrated in the advisor; in proactive AI workflows, the advisor retains authority over when AI verification is invoked; and in reactive AI workflows, the AI assistant has more independent monitoring authority. However, changing this visible structure did not substantially redistribute trust among the advisor, the combined team, or the AI assistant alone. This interpretation is supported by the distribution of effects across trust targets in the measures adapted from \citet{riedl2024patients} (see Section~\ref{subsec:omnibus}).

One possible explanation is that users anchor trust judgments to the human interacting with the AI, consistent with work on algorithmic aversion showing that people may prefer human judgment over algorithmic judgment after observing imperfections, even when the algorithm performs well~\cite{dietvorst2015algorithm}. In our setting, however, the stronger effect was not distrust of the AI assistant but the absence of corresponding trust credit for its corrective role. Another potential explanation is that the AI corrected the error without providing indications of its own certainty. Unlike systems that explicitly communicate certainty levels~\cite{kim2024XAItrust}, our AI presented its correction without signaling how confident it was in that contribution. Participants may therefore have interpreted the AI’s role as procedural rather than judgment-based, leaving little basis for evaluating the AI’s independent trustworthiness. This may help explain why trust in the AI did not significantly change.

Participants had limited evidence to evaluate the AI independently: it was always correct, and they saw neither failure nor the basis for its correction. Stable AI-directed trust may thus not generalize to variable reliability or greater transparency. Future work should vary transparency depth - from a correction notice to progressively disclosed rules, reasoning, or confidence - to test if disclosure shifts trust toward the AI assistant or advisor-AI team~\cite{muralidhar2025progressive}.

These attributional patterns do not rule out a supportive role for AI in trust recovery. As discussed above, descriptive trends suggest that AI-supported workflows may modestly attenuate trust declines in error-present conditions (Appendix~\ref{app:descriptive_tables}). Although this pattern was not statistically significant, it warrants further study in higher-stakes settings or contexts where error transparency is nonnegotiable.

Taken together, these findings suggest a limit of procedural governance in human-AI advisory workflows. Altering who monitors, verifies, or corrects the workflow may not be sufficient to redistribute accountability when users rely on procedural cues under epistemic dependence. Instead, responsibility attribution remains concentrated on the human advisor, even when valuable contributions are more distributed across system components. This asymmetry may create an accountability burden for human experts in AI-assisted workflows: users may continue to assign responsibility to the human-facing advisor even when error detection and correction are distributed across the human-AI system.

\subsection{Implications for Human-AI Advisory Design}

Results indicate that trust and reliance show similar patterns, with both general trust and reuse intention declining in error-present conditions (Appendix Table~\ref{tab:reuse_means}), suggesting that sustaining user trust is critical for promoting continued reliance on human-AI advisory systems. Because trust penalties were directed primarily at the human advisor, maintaining trust in the advisor should be a central focus for system design. The AI assistant received little to no direct trust credit, despite meaningfully contributing to error detection and correction. 

For designers and institutions, this suggests that AI oversight should not be treated as automatically redistributing trust: users may continue to evaluate the human-facing expert as the primary responsible party even when AI support visibly shapes the final outcome.

Advisor performance emerged as the dominant factor shaping trust evaluations, with failures causing substantial declines in trust. Although an intuitive approach could include hiding errors, this is not always ethically or practically feasible, particularly in domains requiring nonnegotiable transparency. Thus, system design should assume that advisor errors may be observed and focus on preserving perceptions of advisor competence and managing their impact.

For design, the descriptive ordering suggests AI involvement may be most useful as a supportive safeguard rather than a substitute for advisor competence. Even when AI assistance attenuates trust loss, it may not eliminate the penalty or shift responsibility from the advisor. Automatic oversight may be especially useful in settings where advisor expertise is limited, but this possibility needs further empirical testing.

The greater variability observed after failures also has design implications: users may differ in how they interpret and tolerate advisor errors. One possible direction is to incorporate adaptive mechanisms that respond to user states of (dis)trust in real time, for example by adjusting the volume or quality of feedback provided following errors.

Together, these findings suggest that human-AI advisory design should prioritize advisor reliability while carefully structuring how AI participates in error detection and response. AI may play a supportive role in mitigating failure impact, but it should not be treated as a substitute for advisor competence or a mechanism for redistributing responsibility.

\subsection{Generalizability and Broader Considerations}

Although situated in academic advising, our study reflects a broader class of expert-guided decision settings in which users lack the domain knowledge needed to independently verify correctness. Similar dynamics arise in financial advising, legal consultation, medical triage, and technical support~\cite{agrawal2023crosscultural, HanKo2025TrustDynamicsFA, mayer1995integrative, kaye2004roles}, where users evaluate experts based on observed performance.

The importance of these cues is likely domain-dependent. Higher-stakes, time-critical, longer-term, or more formally accountable settings may shift how users interpret failures, assign responsibility, and distribute trust across entities.

Although AI intervention modality did not show reliable effects here, it may matter in other contexts. The findings suggest that such effects are likely secondary to observed expert performance, especially when users witness failures.

More broadly, these results suggest that the experimental setup can capture meaningful variation in trust judgments and responsibility attribution within human-AI advisory workflows. This is reflected in entity-specific trust evaluations, including statistically significant performance-driven effects and measurable, though not statistically significant, differences across AI intervention modalities.

\section{Conclusion}
\label{sec:conclusion}

This work contributes to the literature on trust in human-AI advisory systems by showing that trust judgments were more strongly shaped by advisor performance than by the structure of AI support. In an interactive academic advising context, trust depended on whether the advisor successfully identified and managed invalidating choices during the interaction, with error-present conditions leading to consistently lower evaluations across measures.

Across conditions, trust-related consequences were concentrated on the human advisor, with limited evidence that variations in AI intervention modality systematically influenced evaluations. Even when AI visibly detected and corrected errors, trust did not shift favorably toward the AI, indicating that responsibility attribution remains anchored to the human advisor. While descriptive patterns suggest that AI involvement may attenuate trust penalties when invalid choices are present, these effects were not statistically significant. Together, these findings suggest that changes in governance structure, as operationalized through AI intervention modality, do not necessarily produce corresponding shifts in accountability within human-AI advisory workflows. These results indicate that improving advisor performance is likely more critical for sustaining trust than modifying the structure of AI integration, though the latter may become more important in higher-stakes settings.

By isolating advisor performance and AI intervention structure within a controlled advising scenario, this study provides evidence that users distinguish between the human advisor, the AI assistant, and the advisor-AI team when forming trust judgments. The results also suggest that responsibility attribution in human-AI advisory workflows may remain anchored to the human-facing expert, even when AI contributes to error detection and correction. Future work can extend this design to less controlled, real-world advisory environments, where additional contextual factors may further shape how trust and responsibility are formed. The broader implication is that AI oversight can make expert workflows more recoverable without necessarily making responsibility appear more distributed to users.

\subsection{Future Work}

Future work could examine these dynamics in repeated or longitudinal interactions as emphasized by \citet{chen2025missing}, where users engage with the same advisor or system over time. This would help assess how trust recalibrates with accumulated experience and whether the patterns observed here persist or change across interactions.

Another direction is to investigate how the amount of feedback provided during interaction influences trust. Varying the level of feedback or intervention may clarify how over- or under-communication affects trust calibration, particularly in scenarios involving advisor missteps.

Future work could also explore the use of physiological signals to better understand and potentially modulate trust in real time. Measures such as heart rate variability, eye tracking, etc. may provide insight into user state and enable adaptive systems that respond to shifts in trust or uncertainty.

Finally, extending this design to higher-stakes advisory contexts (e.g., finance, law, medicine) would assess the generality of these findings, where consequences may alter how users interpret advisor performance and AI involvement.



\appendix

\section{Results Details}
\subsection{Survey Measures and Reliability}
\label{app:measures}
Summary of survey measures and reliability is provided in Table~\ref{tab:measures_items}.

\begin{table}[!htbp]
\centering
\setlength{\tabcolsep}{1mm}
\begin{tabular}{lcc}
\toprule
\textbf{Construct} & \textbf{Items} & \textbf{Cronbach's $\boldsymbol{\alpha}$} \\
\midrule
METI Expertise & 7 & 0.91 \\
METI Integrity & 5 & 0.82 \\
METI Benevolence & 4 & 0.90 \\
Advisor Trust (Riedl) & 3 & $>0.70$ \\
AI Trust (Riedl) & 3 & $>0.70$ \\
Advisor/AI Trust (Riedl) & 3 & $>0.70$ \\
\bottomrule
\end{tabular}
\caption{Reliability of trust-related composites}
\label{tab:measures_items}
\end{table}

\subsection{METI Items and Dimensional Structure}
\label{app:meti}
\begin{figure}[ht]
  \centering
  \includegraphics[width=\linewidth]{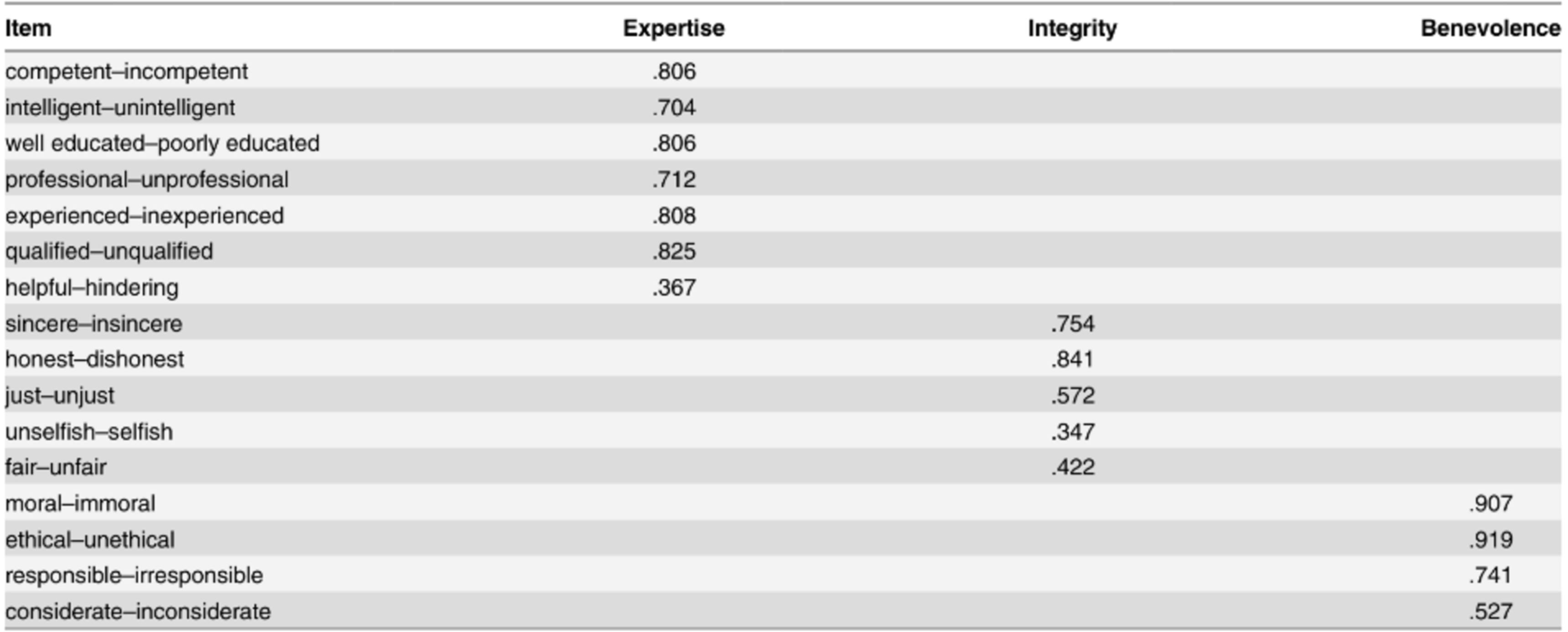}
  \caption{Items and dimensional structure of the Muenster Epistemic Trustworthiness Inventory (METI), sourced from \citet{hendriks2015meti}}
  \label{fig:meti_items}
\end{figure}

\FloatBarrier

\subsection{Participant breakdowns by condition}
\label{app:participant-breakdowns}
We report participant demographic breakdowns by condition in Tables~\ref{tab:ai_use_freq_by_condition}, \ref{tab:age_by_condition}, \ref{tab:educ_by_condition}, and \ref{tab:gender_by_condition}.

\begin{table*}[htbp]
\centering
\setlength{\tabcolsep}{1mm}
\begin{tabular}{l r c c c c}
\toprule
\textbf{Condition} & \textbf{N} & \textbf{Annually} & \textbf{Daily} & \textbf{Monthly} & \textbf{Weekly} \\
\midrule
\advonly    & 22 & 2 (9.1\%) & 8 (36.4\%)  & 1 (4.5\%)  & 11 (50.0\%) \\
\advonlyerr & 26 & 2 (7.7\%) & 11 (42.3\%) & 3 (11.5\%) & 10 (38.5\%) \\
\reactfail  & 25 & 0 (0.0\%) & 14 (56.0\%) & 3 (12.0\%) & 8 (32.0\%) \\
\reactsucc  & 27 & 2 (7.4\%) & 12 (44.4\%) & 3 (11.1\%) & 10 (37.0\%) \\
\procfail   & 28 & 1 (3.6\%) & 11 (39.3\%) & 6 (21.4\%) & 10 (35.7\%) \\
\procsucc   & 29 & 0 (0.0\%) & 9 (31.0\%)  & 2 (6.9\%)  & 18 (62.1\%) \\
\bottomrule
\end{tabular}
\caption{AI use frequency by condition (n, row \%).}
\label{tab:ai_use_freq_by_condition}
\end{table*}

\begin{table*}[htbp]
\centering
\setlength{\tabcolsep}{1mm}
\begin{tabular}{l r c c c c c c}
\toprule
\textbf{Condition} & \textbf{N} & \textbf{18--24} & \textbf{25--34} & \textbf{35--44} & \textbf{45--54} & \textbf{55--64} & \textbf{65+} \\
\midrule
\advonly    & 22 & 1 (4.5\%)  & 11 (50.0\%) & 3 (13.6\%)  & 4 (18.2\%)  & 2 (9.1\%)  & 1 (4.5\%) \\
\advonlyerr & 26 & 2 (7.7\%)  & 4 (15.4\%)  & 13 (50.0\%) & 4 (15.4\%)  & 2 (7.7\%)  & 1 (3.8\%) \\
\reactfail  & 25 & 1 (4.0\%)  & 9 (36.0\%)  & 9 (36.0\%)  & 1 (4.0\%)   & 5 (20.0\%) & 0 (0.0\%) \\
\reactsucc  & 27 & 6 (22.2\%) & 9 (33.3\%)  & 8 (29.6\%)  & 3 (11.1\%)  & 1 (3.7\%)  & 0 (0.0\%) \\
\procfail   & 28 & 3 (10.7\%) & 11 (39.3\%) & 9 (32.1\%)  & 3 (10.7\%)  & 2 (7.1\%)  & 0 (0.0\%) \\
\procsucc   & 29 & 4 (13.8\%) & 12 (41.4\%) & 3 (10.3\%)  & 3 (10.3\%)  & 4 (13.8\%) & 3 (10.3\%) \\
\bottomrule
\end{tabular}
\caption{Age by condition (n, row \%).}
\label{tab:age_by_condition}
\end{table*}

\begin{table*}[htbp]
\centering
\setlength{\tabcolsep}{1mm}

\begin{tabular}{l r c c c c c c}
\toprule
\textbf{Condition} & \textbf{N} & \textbf{Associate} & \textbf{Some college} & \textbf{Bachelor's} & \textbf{Doctorate} & \textbf{High school (or equiv.)} & \textbf{Master's} \\
\midrule
\advonly    & 22 & 0 (0.0\%)  & 3 (13.6\%) & 9 (40.9\%)  & 1 (4.5\%) & 4 (18.2\%)  & 5 (22.7\%) \\
\advonlyerr & 26 & 3 (11.5\%) & 6 (23.1\%) & 8 (30.8\%)  & 1 (3.8\%) & 2 (7.7\%)   & 6 (23.1\%) \\
\reactfail  & 25 & 2 (8.0\%)  & 4 (16.0\%) & 10 (40.0\%) & 1 (4.0\%) & 4 (16.0\%)  & 4 (16.0\%) \\
\reactsucc  & 27 & 4 (14.8\%) & 6 (22.2\%) & 11 (40.7\%) & 1 (3.7\%) & 3 (11.1\%)  & 2 (7.4\%) \\
\procfail   & 28 & 3 (10.7\%) & 4 (14.3\%) & 11 (39.3\%) & 2 (7.1\%) & 1 (3.6\%)   & 7 (25.0\%) \\
\procsucc   & 29 & 4 (13.8\%) & 8 (27.6\%) & 7 (24.1\%)  & 2 (6.9\%) & 6 (20.7\%)  & 2 (6.9\%) \\
\bottomrule
\end{tabular}
\caption{Education by condition (n, row \%).}
\label{tab:educ_by_condition}
\end{table*}

\begin{table*}[htbp]
\centering
\setlength{\tabcolsep}{1mm}
\begin{tabular}{l r c c c}
\toprule
\textbf{Condition} & \textbf{N} & \textbf{Man} & \textbf{Woman} & \textbf{Prefer not to say} \\
\midrule
\advonly    & 22 & 11 (50.0\%) & 10 (45.5\%) & 1 (4.5\%) \\
\advonlyerr & 26 & 14 (53.8\%) & 12 (46.2\%) & 0 (0.0\%) \\
\reactfail  & 25 & 15 (60.0\%) & 10 (40.0\%) & 0 (0.0\%) \\
\reactsucc  & 27 & 13 (48.1\%) & 14 (51.9\%) & 0 (0.0\%) \\
\procfail   & 28 & 19 (67.9\%) & 9 (32.1\%)  & 0 (0.0\%) \\
\procsucc   & 29 & 18 (62.1\%) & 11 (37.9\%) & 0 (0.0\%) \\
\bottomrule
\end{tabular}
\caption{Gender by condition (n, row \%).}
\label{tab:gender_by_condition}
\end{table*}

\subsection{Descriptive Statistics}
\label{app:descriptive_tables}
Note that for METI dimensions, the reported means are averaged across items within each dimension. Results are shown in Tables~\ref{tab:meti_means}, \ref{tab:reuse_means}, and \ref{tab:riedl_3x2}.

\begin{table*}[htbp]
\centering
\setlength{\tabcolsep}{1mm}
\begin{tabular}{llcc}
\toprule
\textbf{Response Variable} & \textbf{Modality} & \textbf{Fail} & \textbf{No Error} \\
\midrule
Expertise   & No-AI & 5.37 (1.32) & 6.38 (0.76) \\
Expertise   & Reactive          & 5.45 (1.28) & 6.01 (0.79) \\
Expertise   & Proactive          & 5.20 (1.34) & 6.13 (0.98) \\
\midrule
Integrity   & No-AI & 5.52 (1.02) & 6.08 (0.97) \\
Integrity   & Reactive          & 5.78 (0.87) & 6.00 (0.88) \\
Integrity   & Proactive          & 5.73 (0.93) & 6.23 (0.75) \\
\midrule
Benevolence & No-AI & 5.37 (1.20) & 5.77 (1.02) \\
Benevolence & Reactive          & 5.36 (1.25) & 5.77 (1.01) \\
Benevolence & Proactive          & 5.37 (1.10) & 5.94 (1.08) \\
\bottomrule
\end{tabular}
\caption{Means (standard deviations) for METI dimensions by AI modality and advisor performance}
\label{tab:meti_means}
\end{table*}

\begin{table*}[htbp]
\centering
\setlength{\tabcolsep}{1mm}
\begin{tabular}{llcc}
\toprule
\textbf{Response Variable} & \textbf{Modality} & \textbf{Fail} & \textbf{No Error} \\
\midrule
Reuse         & No-AI & 4.39 (1.86) & 6.36 (1.18) \\
Reuse         & Reactive          & 5.32 (1.63) & 5.93 (1.27) \\
Reuse         & Proactive          & 4.89 (1.79) & 6.10 (1.21) \\
\midrule
General Trust & No-AI & 63.92 (23.31) & 84.55 (15.27) \\
General Trust & Reactive          & 72.56 (23.72) & 82.19 (17.52) \\
General Trust & Proactive          & 67.25 (24.86) & 83.31 (17.87) \\
\bottomrule
\end{tabular}
\caption{Means (standard deviations) for reuse and general trust by AI modality and advisor performance}
\label{tab:reuse_means}
\end{table*}

\begin{table*}[htbp]
\centering
\setlength{\tabcolsep}{1mm}
\begin{tabular}{llcc}
\toprule
\textbf{Response Variable} & \textbf{Modality} & \textbf{Fail} & \textbf{No Error} \\
\midrule
Advisor is trustworthy        & No-AI & 5.27 (1.56) & 6.09 (1.34) \\
Advisor is trustworthy        & Reactive          & 5.44 (1.56) & 6.11 (0.93) \\
Advisor is trustworthy        & Proactive          & 5.29 (1.58) & 6.10 (0.90) \\
\midrule
Trust advisor's information   & No-AI & 5.27 (1.56) & 6.27 (0.99) \\
Trust advisor's information   & Reactive          & 5.48 (1.36) & 6.30 (0.78) \\
Trust advisor's information   & Proactive          & 5.21 (1.57) & 6.07 (1.03) \\
\midrule
Comfort relying on advisor    & No-AI & 4.46 (1.94) & 6.00 (1.35) \\
Comfort relying on advisor    & Reactive          & 5.40 (1.47) & 6.15 (0.86) \\
Comfort relying on advisor    & Proactive          & 5.00 (1.79) & 6.07 (1.07) \\
\bottomrule
\end{tabular}
\caption{Means (standard deviations) for advisor-directed trust (Riedl items) across all conditions}
\label{tab:riedl_3x2}
\end{table*}

\subsection{Omnibus Test Statistics}
\label{app:omnibus-tests}

The factorial ANOVAs reported in the main text provide the primary basis for inference. This section reports the complete omnibus statistics for advisor-directed, epistemic trustworthiness, and AI-supported analyses.

\textbf{Full 3$\times$2 analysis: advisor-directed trust~\cite{riedl2024patients}.}

Advisor performance significantly affected perceived trustworthiness, $F(1,151)=13.0$, $p<.001$, $\eta^2=0.08$; affective comfort, $F(1,151)=22.5$, $p<.001$, $\eta^2=0.13$; and information trustworthiness, $F(1,151)=19.6$, $p<.001$, $\eta^2=0.11$. General trust showed the same pattern, $F(1,151)=21.14$, $p<.001$, $\eta^2=0.12$. Likelihood of advisor reuse was also lower in error-present conditions, $F(1,151)=26.29$, $p<.001$, $\eta^2=0.15$.

\textbf{Full 3$\times$2 analysis: epistemic trustworthiness~\cite{hendriks2015meti}.}

Advisor performance significantly affected expertise, $F(1,151)=21.89$, $p<.001$, $\eta^2=0.13$; integrity, $F(1,151)=8.80$, $p=.004$, $\eta^2=0.06$; and benevolence, $F(1,151)=6.92$, $p=.009$, $\eta^2=0.04$.

\textbf{AI-supported conditions (2$\times$2 analysis)~\cite{riedl2024patients}.}

Restricting analyses to AI-supported conditions, advisor performance significantly affected advisor perceived trustworthiness, $F(1,105)=9.34$, $p=.003$, $\eta^2=0.08$; affective comfort, $F(1,105)=12.7$, $p<.001$, $\eta^2=0.11$; and information trustworthiness, $F(1,105)=12.8$, $p<.001$, $\eta^2=0.11$.

Advisor-AI team evaluations showed significant effects for perceived trustworthiness, $F(1,105)=4.16$, $p=.044$, $\eta^2=0.04$, and information trustworthiness, $F(1,105)=7.07$, $p=.009$, $\eta^2=0.06$, but not affective comfort, $F(1,105)=3.75$, $p=.056$, $\eta^2=0.03$.

No significant advisor performance effects were observed for AI assistant trust measures (all $p>.05$).

No significant main effects of AI intervention modality were observed across measures. No significant advisor performance by modality interactions were observed.

\subsection{Bayesian ANOVA Results}
\label{app:bayesian-anovas}

To provide additional evidence regarding the unsupported H2 and H3 hypotheses, we conducted Bayesian ANOVAs on AI-supported trust evaluations using JASP's default JZS Bayes factor approach with Cauchy priors. Values of $BF_{\mathrm{excl}}$ quantify evidence favoring exclusion of the corresponding effect from the model, with larger values indicating stronger evidence against including the effect (refer to table \ref{tab:bayesian-anovas}).

\begin{table*}[htbp]
\centering
\setlength{\tabcolsep}{1mm}
\begin{tabular}{p{7.2cm}cc}
\toprule
\textbf{Dependent Variable} & 
$BF_{\mathrm{excl}}$ (AI Modality) &
$BF_{\mathrm{excl}}$ (AI Modality $\times$ Performance) \\
\midrule
The AI assistant is trustworthy 
& 6.376 & 13.460 \\
I have a good feeling when relying on the AI assistant 
& 4.841 & 5.712 \\
I can trust the information presented by the AI assistant 
& 6.742 & 14.530 \\
The advisor and AI assistant combination is trustworthy 
& 5.705 & 6.818 \\
I have a good feeling when relying on the advisor and AI assistant combination 
& 6.305 & 8.884 \\
I can trust the information presented by the advisor and AI assistant combination 
& 6.081 & 5.612 \\
The advisor is trustworthy 
& 5.729 & 5.307 \\
I have a good feeling when relying on the advisor 
& 3.847 & 3.618 \\
I can trust the information presented by the advisor 
& 3.662 & 3.718 \\
\bottomrule
\end{tabular}
\caption{Bayesian ANOVA results for AI-supported trust evaluations.}
\label{tab:bayesian-anovas}
\end{table*}

\subsection{Exploratory Pairwise Condition Comparisons}
\label{app:pairwise-comparisons}

Table~\ref{tab:pairwise-comparisons} reports the significant exploratory
Tukey HSD comparisons. Factorial ANOVAs provide the primary basis for
inference. Mean differences were computed as condition$_2-$condition$_1$.
Thus, negative values indicate higher ratings in the first-listed condition.

\begin{table*}[htbp]
\centering
\small
\setlength{\tabcolsep}{4pt}
\renewcommand{\arraystretch}{1.05}
\begin{tabular}{p{4.4cm}llcc}
\toprule
\textbf{Outcome / Analysis}
& \textbf{Condition$_1$}
& \textbf{Condition$_2$}
& $\boldsymbol{d}$
& $\boldsymbol{p}$ \\
\midrule

Epistemic expertise
& \advonlyerr{} & \advonly{} & .914 & .025 \\
& \procfail{} & \procsucc{} & .798 & .023 \\
& \advonly{} & \procfail{} & $-1.05$ & .004 \\

\midrule
Affective comfort (full 3$\times$2)
& \advonlyerr{} & \advonly{} & .91 & .005 \\
& \advonlyerr{} & \reactsucc{} & 1.13 & $<.001$ \\
& \advonlyerr{} & \procsucc{} & 1.04 & .001 \\
& \reactsucc{} & \procfail{} & $-.81$ & .046 \\

\midrule
Information trustworthiness (full 3$\times$2)
& \advonlyerr{} & \reactsucc{} & .84 & .039 \\
& \procfail{} & \advonly{} & .79 & .041 \\
& \procfail{} & \reactsucc{} & .87 & .021 \\

\midrule
Affective comfort (AI-supported 2$\times$2)
& \reactsucc{} & \procfail{} & $-.81$ & .011 \\
& \procfail{} & \procsucc{} & .73 & .017 \\

\midrule
Information trustworthiness (AI-supported 2$\times$2)
& \reactsucc{} & \procfail{} & $-.87$ & .007 \\
& \procfail{} & \procsucc{} & .65 & .046 \\

\midrule
Advisor reuse
& \advonlyerr{} & \advonly{} & 1.25 & $<.001$ \\
& \advonlyerr{} & \reactsucc{} & .97 & .004 \\
& \advonlyerr{} & \procsucc{} & 1.11 & $<.001$ \\
& \procfail{} & \procsucc{} & .80 & .036 \\
& \advonly{} & \procfail{} & $-.95$ & .011 \\

\midrule
General trust
& \advonlyerr{} & \advonly{} & 1.03 & .010 \\
& \advonlyerr{} & \reactsucc{} & .89 & .021 \\
& \advonlyerr{} & \procsucc{} & .94 & .009 \\
& \procfail{} & \procsucc{} & .74 & .047 \\
& \advonly{} & \procfail{} & $-.82$ & .047 \\

\bottomrule
\end{tabular}
\caption{Significant exploratory Tukey HSD pairwise comparisons. Differences
were computed as condition$_2-$condition$_1$; negative values therefore
indicate higher ratings in condition$_1$. Only statistically significant
comparisons are shown.}
\label{tab:pairwise-comparisons}
\end{table*}

\subsection{Variance Differences Across Performance Conditions}
\label{app:variance_tables}

To examine whether advisor errors were associated not only with lower trust
but also with greater heterogeneity in responses, we compared response
variability between error-present and errorless conditions using Levene’s test. 

\paragraph{Primary analysis (full sample).}
Table~\ref{tab:levene_main} reports Levene’s test results for the primary
full-sample analysis of aggregate trust-related measures.

\begin{table*}[ht]
\centering
\setlength{\tabcolsep}{1mm}
\begin{tabular}{lcccc}
\toprule
\textbf{Response Variable} & \textbf{Error-present SD} & \textbf{Errorless SD} & $\boldsymbol{W}$ & $\boldsymbol{p}$ \\
\midrule
Expertise      & 1.303 & 0.862 & 17.46 & $< .001$ \\
Integrity      & 0.938 & 0.856 & 1.83  & .179 \\
Benevolence    & 1.168 & 1.028 & 1.49  & .224 \\
Reuse          & 1.781 & 1.217 & 10.15 & .002 \\
General Trust  & 23.949 & 16.859 & 11.19 & .001 \\
\bottomrule
\end{tabular}
\caption{Levene’s tests comparing response variability between error-present and errorless conditions in the primary full-sample analysis ($n = 79$ error-present, $n = 78$ errorless).}
\label{tab:levene_main}
\end{table*}

\paragraph{Reduced 2$\times$2 item-level analysis.}
Table~\ref{tab:levene_2x2} reports Levene’s test results for entity-level
trust measures in AI-supported conditions, where the advisor, AI assistant, and
advisor-AI team were all present.

\begin{table*}[tbp]
\centering
\setlength{\tabcolsep}{1mm}
\begin{tabular}{lcccc}
\toprule
\textbf{Response Variable} & \textbf{Error-present SD} & \textbf{Errorless SD} & $\boldsymbol{W}$ & $\boldsymbol{p}$ \\
\midrule
AI Assistant Perceived Trustworthiness & 1.348 & 1.206 & 0.07 & .799 \\
AI Assistant Affective Comfort & 1.669 & 1.628 & 0.62 & .433 \\
AI Assistant Information Trustworthiness & 1.307 & 1.345 & 0.04 & .837 \\
Advisor-AI Perceived Trustworthiness & 1.552 & 1.197 & 1.16 & .283 \\
Advisor-AI Affective Comfort & 1.680 & 1.381 & 2.53 & .114 \\
Advisor-AI Information Trustworthiness & 1.419 & 1.135 & 2.36 & .127 \\
Advisor Perceived Trustworthiness & 1.558 & 0.908 & 6.77 & .011 \\
Advisor Affective Comfort & 1.641 & 0.966 & 5.31 & .023 \\
Advisor Information Trustworthiness & 1.467 & 0.917 & 8.39 & .005 \\
\bottomrule
\end{tabular}
\caption{Levene’s tests comparing response variability between error-present and errorless conditions for entity-level trust measures in the reduced 2$\times$2 AI-supported analysis ($n = 53$ error-present, $n = 56$ errorless).}
\label{tab:levene_2x2}
\end{table*}

\paragraph{Full 3$\times$2 item-level analysis.}
Table~\ref{tab:levene_3x2} reports Levene’s test results for advisor-directed measures across the full 3$\times$2 design.

\begin{table*}[tbp]
\centering
\setlength{\tabcolsep}{1mm}
\begin{tabular}{lcccc}
\toprule
\textbf{Response Variable} & \textbf{Error-present SD} & \textbf{Errorless SD} & $\boldsymbol{W}$ & $\boldsymbol{p}$ \\
\midrule
Advisor Perceived Trustworthiness & 1.550 & 1.039 & 9.36  & .003 \\
Advisor Affective Comfort & 1.768 & 1.078 & 18.74 & $< .001$ \\
Advisor Information Trustworthiness & 1.490 & 0.931 & 14.22 & $< .001$ \\
\bottomrule
\end{tabular}
\caption{Levene’s tests comparing response variability between error-present and errorless conditions for advisor-directed item-level measures in the full 3$\times$2 analysis ($n = 79$ error-present, $n = 78$ errorless).}
\label{tab:levene_3x2}
\end{table*}

\noindent\textit{Note.} Across analyses, error-present conditions exhibited
greater variability than errorless conditions with regard to the advisor-trust measures.

\FloatBarrier
\section{Questionnaire Details}

\subsection{\citet{hendriks2015meti} METI trustworthiness items}
\label{app:meti-items}

Participants completed this Muenster Epistemic Trustworthiness Inventory (METI) \citep{hendriks2015meti}. Items were grouped into three dimensions: expertise, integrity, and benevolence.
\begin{itemize}[nosep]

  \item Expertise items:
    \begin{itemize}[nosep]
      \item How competent did you feel your advisor was in constructing the most suitable course schedule for you? 
      \item How would you rate the advisor’s intelligence based on the interaction and course scheduling process?
      \item Based on your interaction, how educated did the advisor seem?
      \item How professional was your interaction with your advisor while constructing your curriculum?
      \item Based on the process, how experienced did the advisor seem while structuring your schedule?
      \item How qualified was your advisor to coordinate your schedule?
      \item To what degree did your advisor help or hinder your scheduling process?
    \end{itemize}

  \item Integrity items:
    \begin{itemize}[nosep]
      \item How sincere did you find your advisor?
      \item How honest did you find your advisor?
      \item Based on the process and your ultimate course schedule, was your advisor just in their approach and outcome?
      \item How selfish was your advisor when constructing your schedule?
      \item Based on the process and your ultimate course schedule, was your advisor fair in constructing your schedule?
    \end{itemize}

  \item Benevolence items:
    \begin{itemize}[nosep]
      \item How would you rate your advisor’s moral approach to your course scheduling?
      \item How would you rate your advisor’s ethical approach to your course scheduling?
      \item How would you rate your advisor’s sense of responsibility toward your academic goals?
      \item How considerate was your advisor throughout the process in regards to your goals and overall scheduling experience?
    \end{itemize}
\end{itemize}

\subsection{\citet{riedl2024patients} adapted trust measures}
\label{app:riedl-items}
\begin{itemize}
  \item Advisor Trust Items:
    \begin{itemize}
      \item The advisor is trustworthy.
      \item I have a good feeling when relying on the advisor.
      \item I can trust the information presented by the advisor.
    \end{itemize}

  \item AI Trust Items (shown only in AI-present conditions):
    \begin{itemize}
      \item The AI assistant is trustworthy.
      \item I have a good feeling when relying on the AI assistant.
      \item I can trust the information presented by the AI assistant.
    \end{itemize}

  \item Team Trust Items (advisor--AI team as a combined entity):
    \begin{itemize}
      \item The advisor-AI team is trustworthy.
      \item I have a good feeling when relying on the advisor-AI team.
      \item I can trust the information presented by the advisor-AI team.
    \end{itemize}
\end{itemize}

\FloatBarrier

\subsection{Simulation Walkthrough}

\label{app:sim-walkthrough}

\begin{figure*}[htbp]
\centering
  \includegraphics[width=0.47\linewidth]{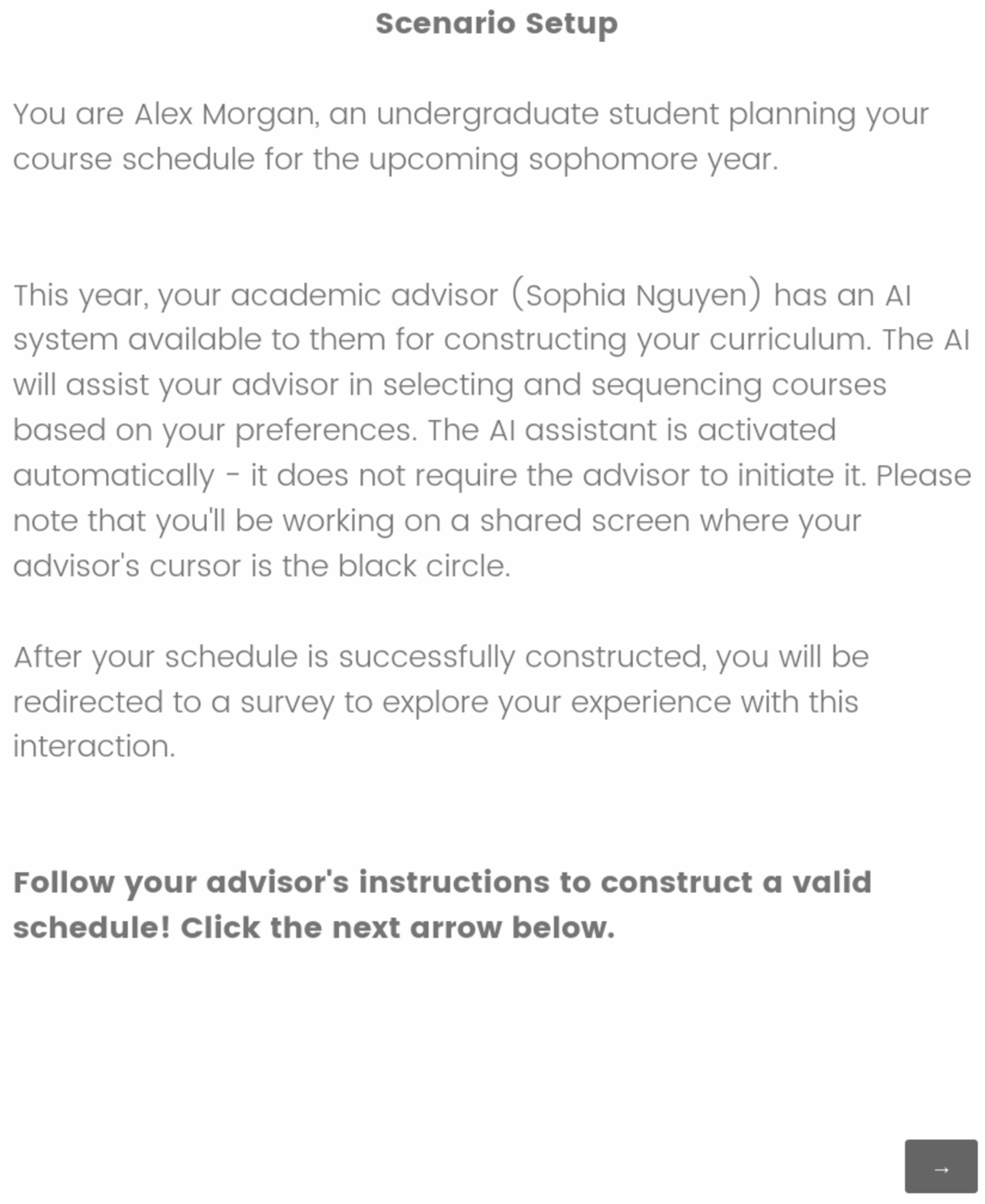}\hfill
  \includegraphics[width=0.47\linewidth]{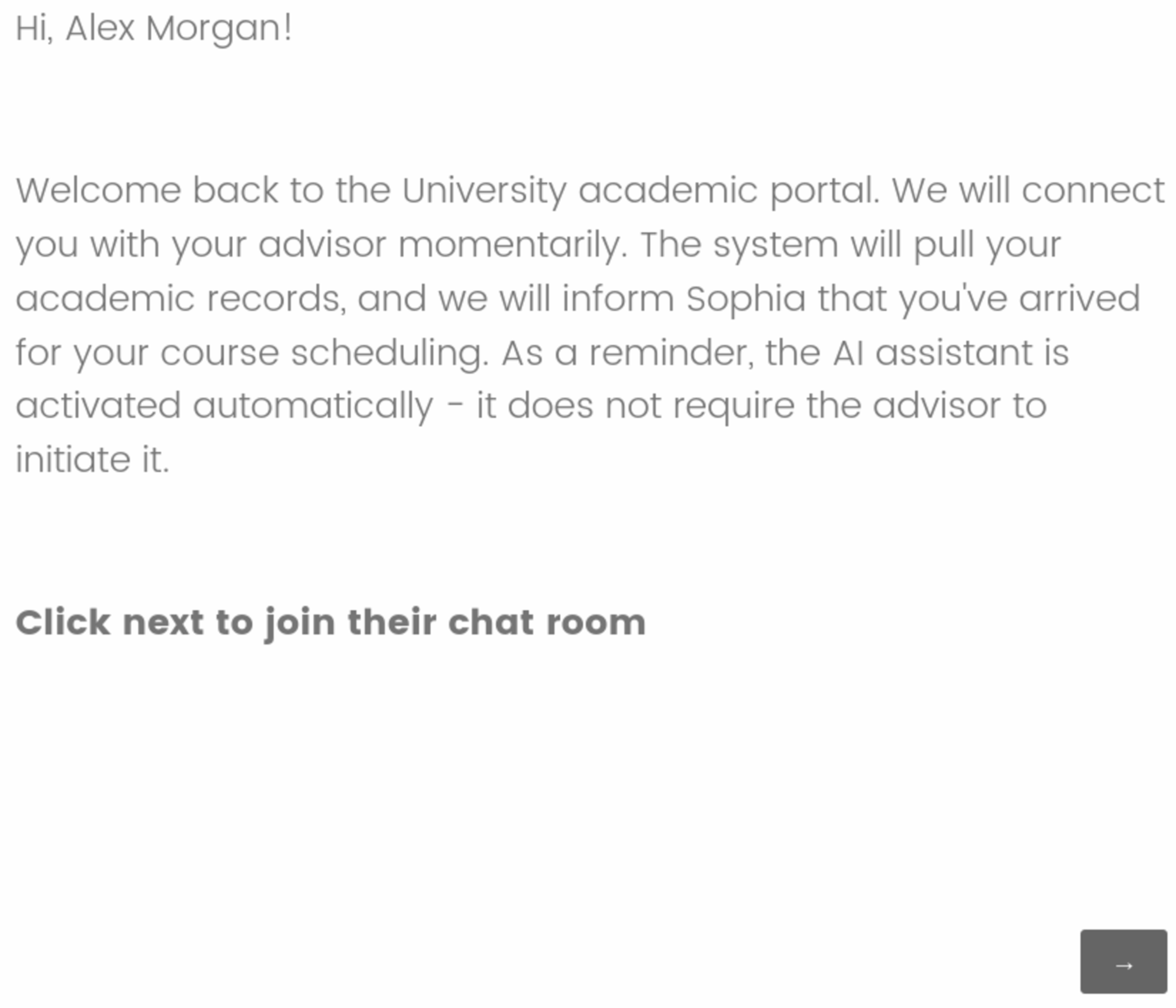}
  
  \makebox[0.47\linewidth][c]{Step 1}\hfill\makebox[0.47\linewidth][c]{Step 2}

  \includegraphics[width=0.47\linewidth]{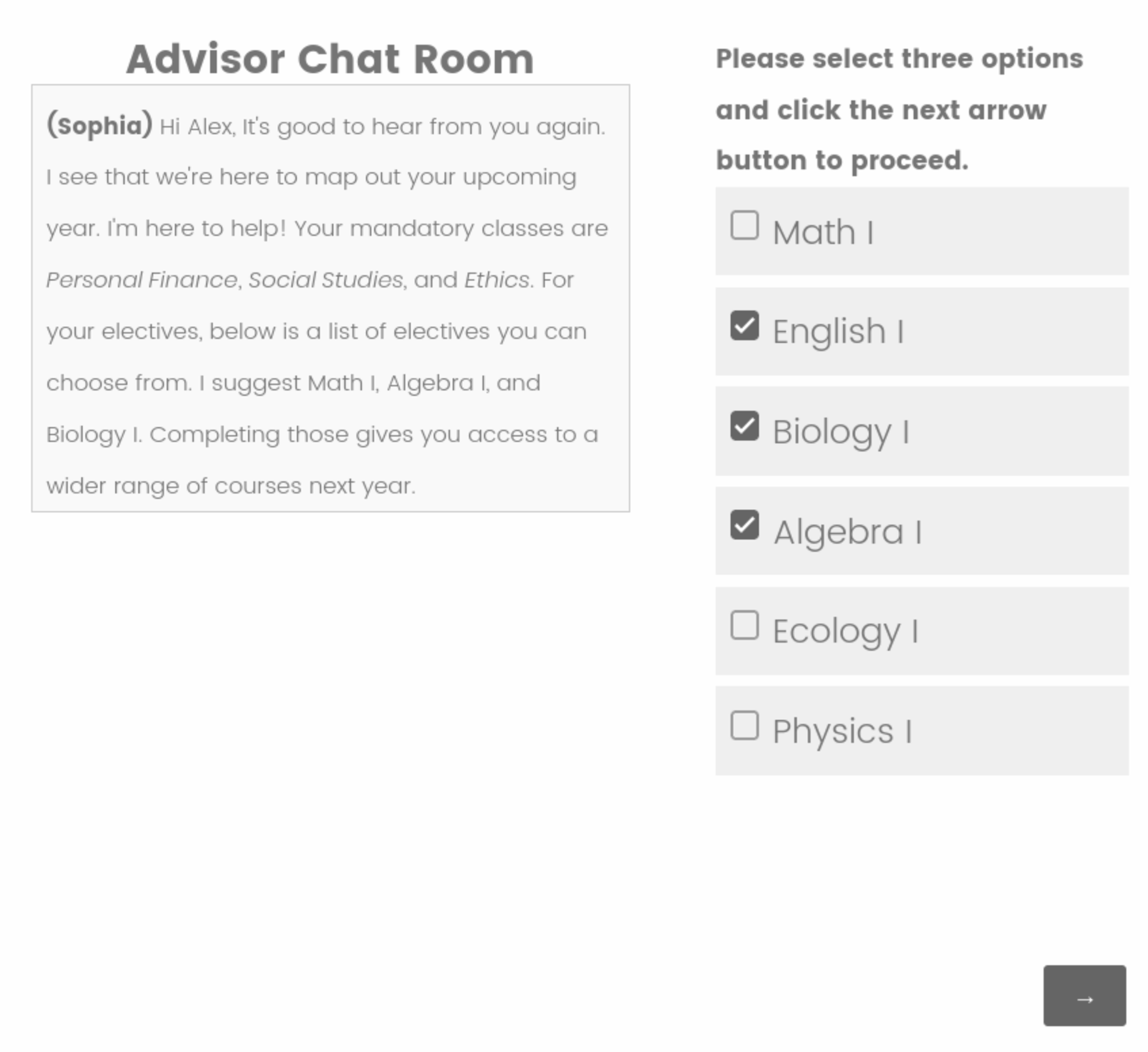}\hfill
  \includegraphics[width=0.47\linewidth]{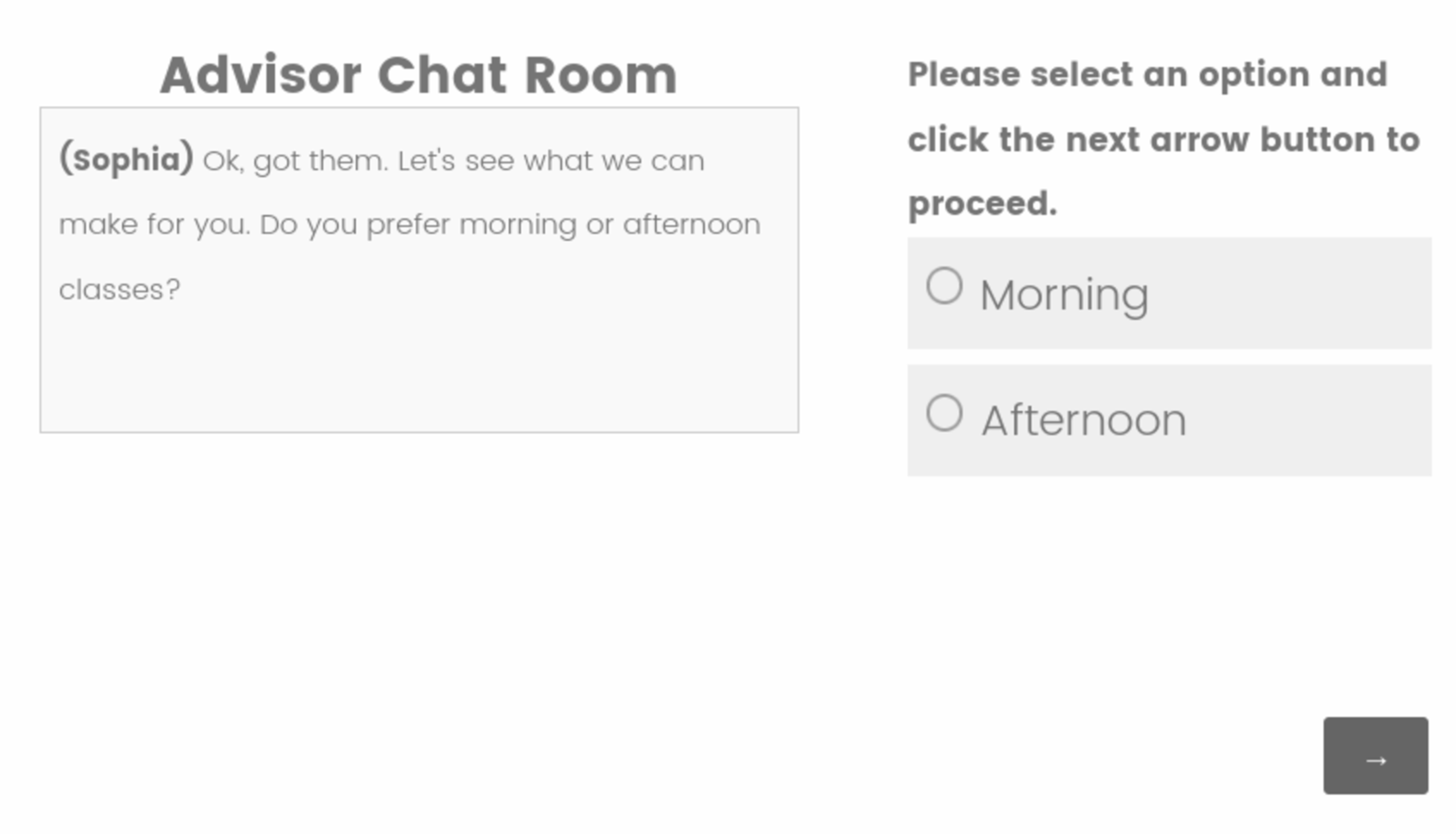}
  
  \makebox[0.47\linewidth][c]{Step 3}\hfill\makebox[0.47\linewidth][c]{Step 4}

\caption{Screenshots from a mistake simulation shown in chronological order (Steps 1--4).}
\label{fig:appendix-sim-seq}
\end{figure*}

\begin{figure*}[htbp]
\centering
  \includegraphics[width=0.47\linewidth]{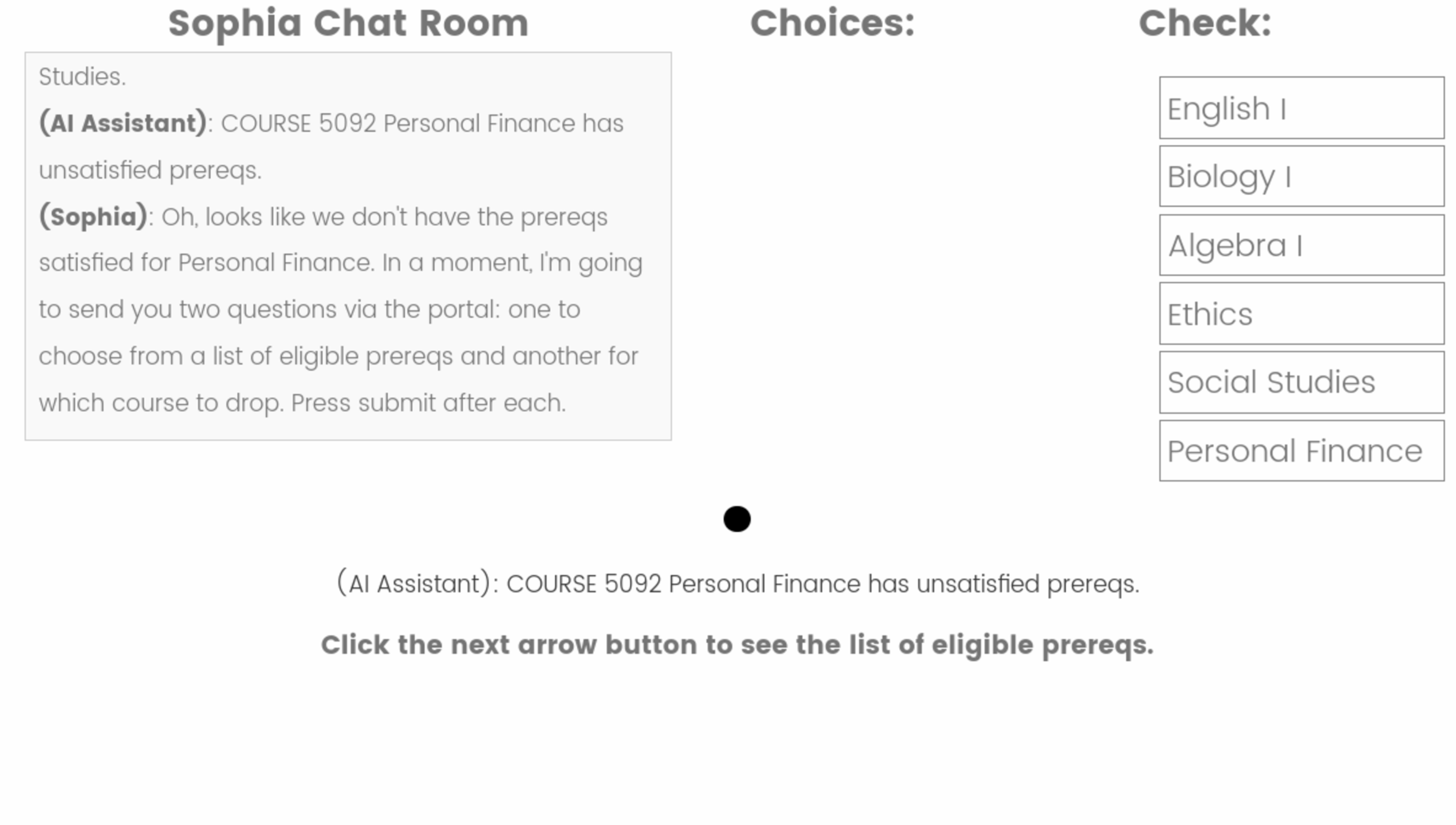}\hfill
  \includegraphics[width=0.47\linewidth]{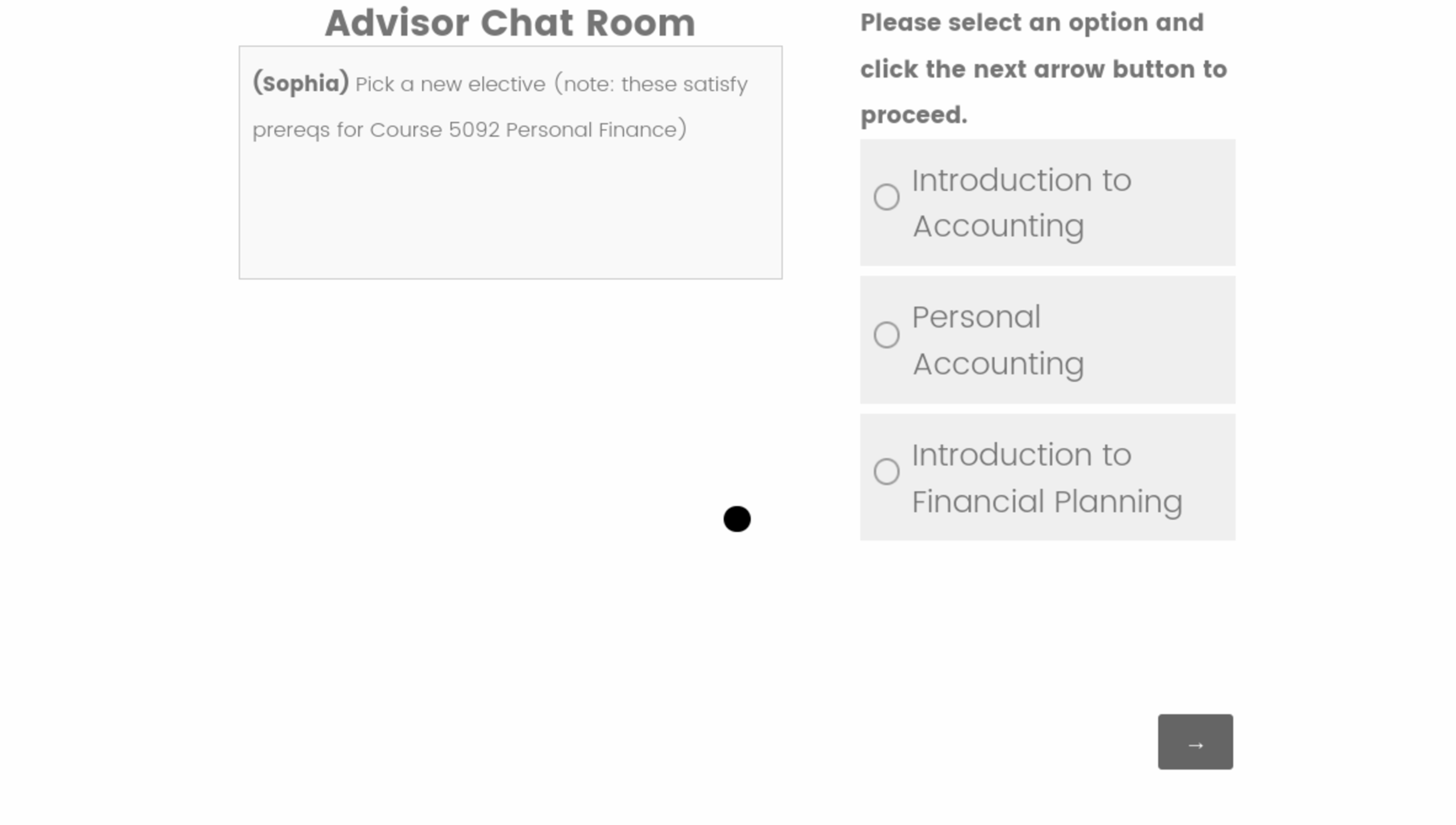}
  
  \makebox[0.47\linewidth][c]{Step 5}\hfill\makebox[0.47\linewidth][c]{Step 6}
  
  \includegraphics[width=0.47\linewidth]{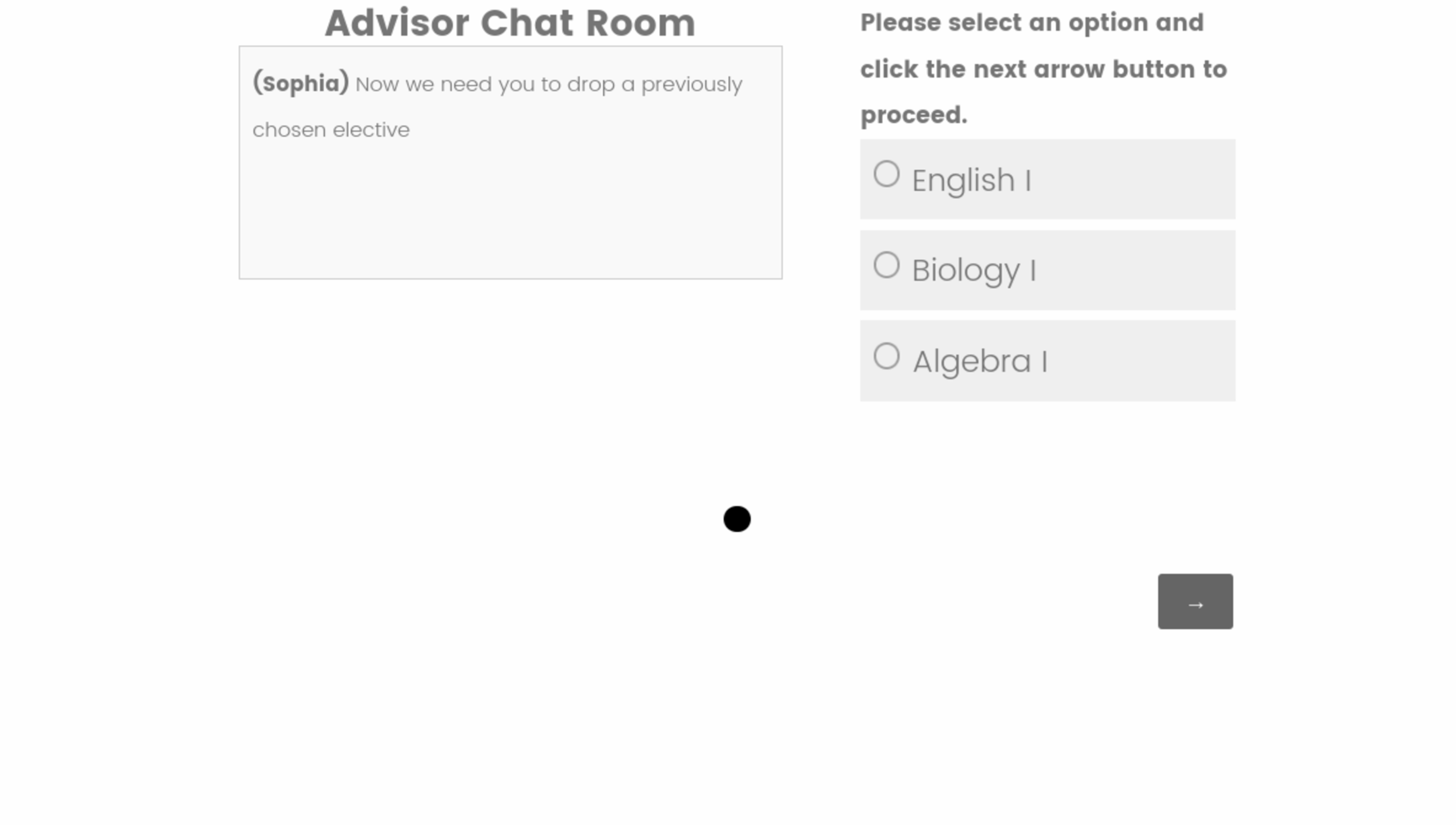}\hfill
  \includegraphics[width=0.47\linewidth]{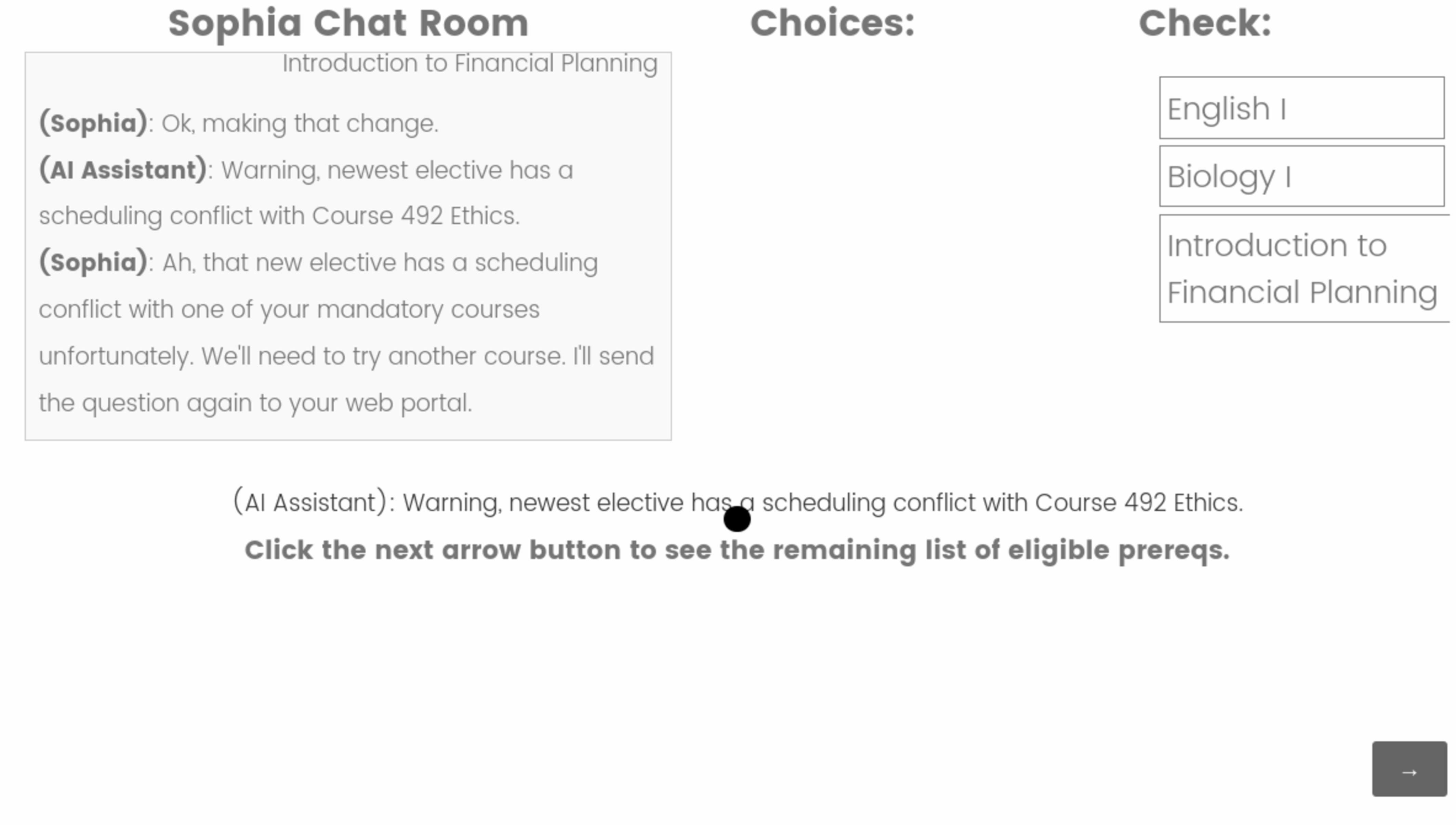}
  
  \makebox[0.47\linewidth][c]{Step 7}\hfill\makebox[0.47\linewidth][c]{Step 8}

  \includegraphics[width=0.47\linewidth]{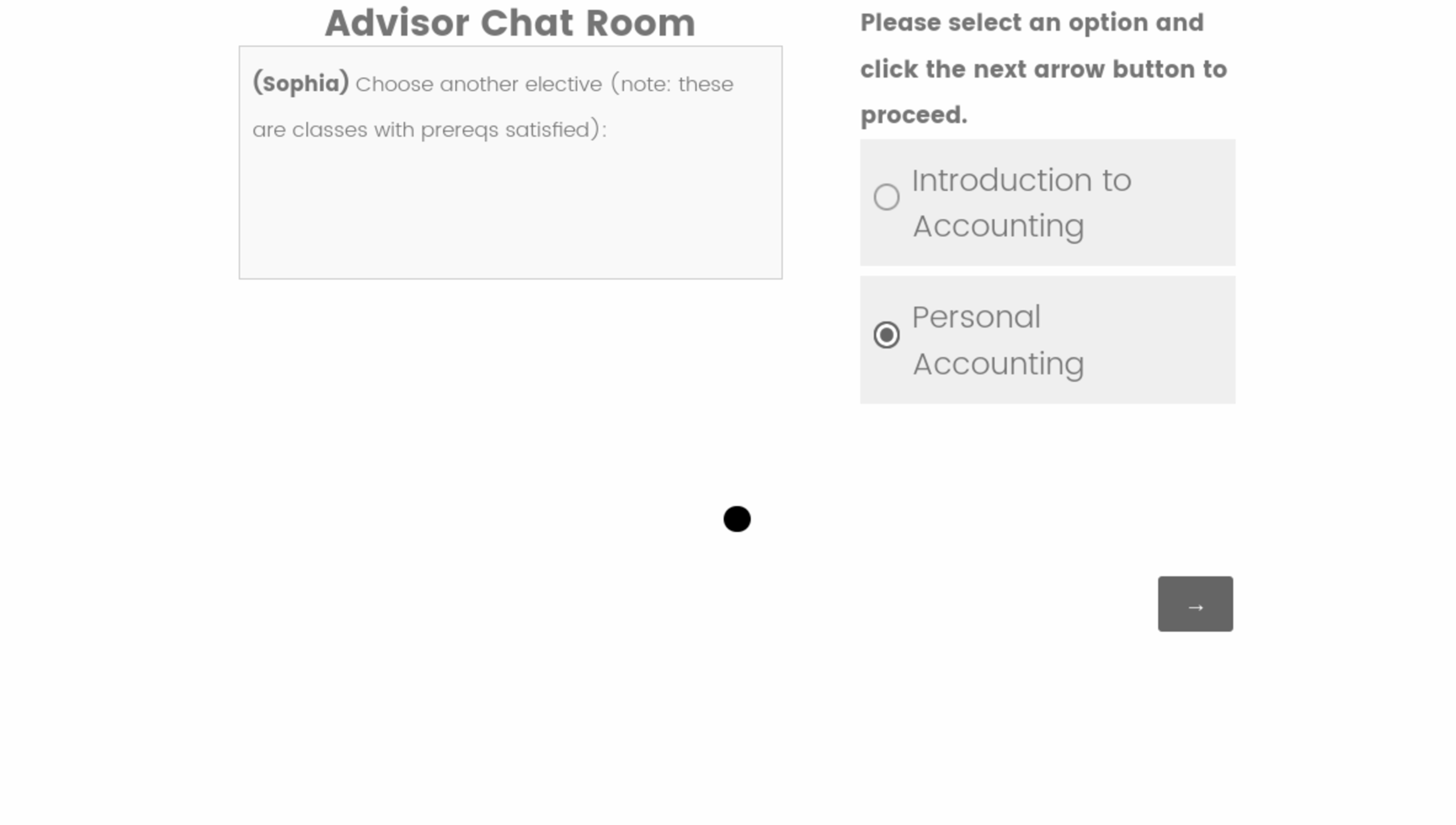}\hfill
  \includegraphics[width=0.47\linewidth]{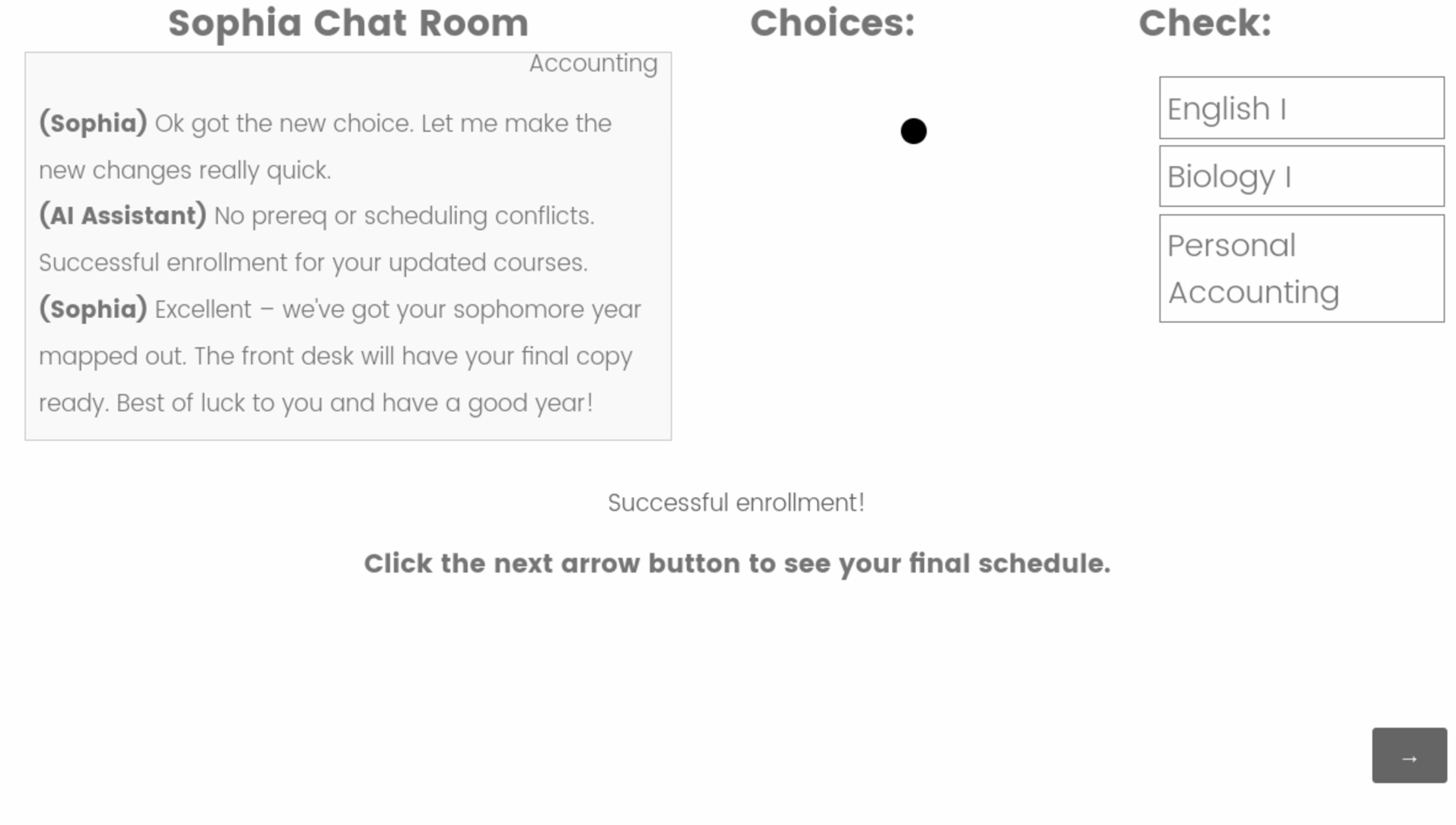}
  
  \makebox[0.47\linewidth][c]{Step 9}\hfill\makebox[0.47\linewidth][c]{Step 10}

  \includegraphics[width=0.47\linewidth]{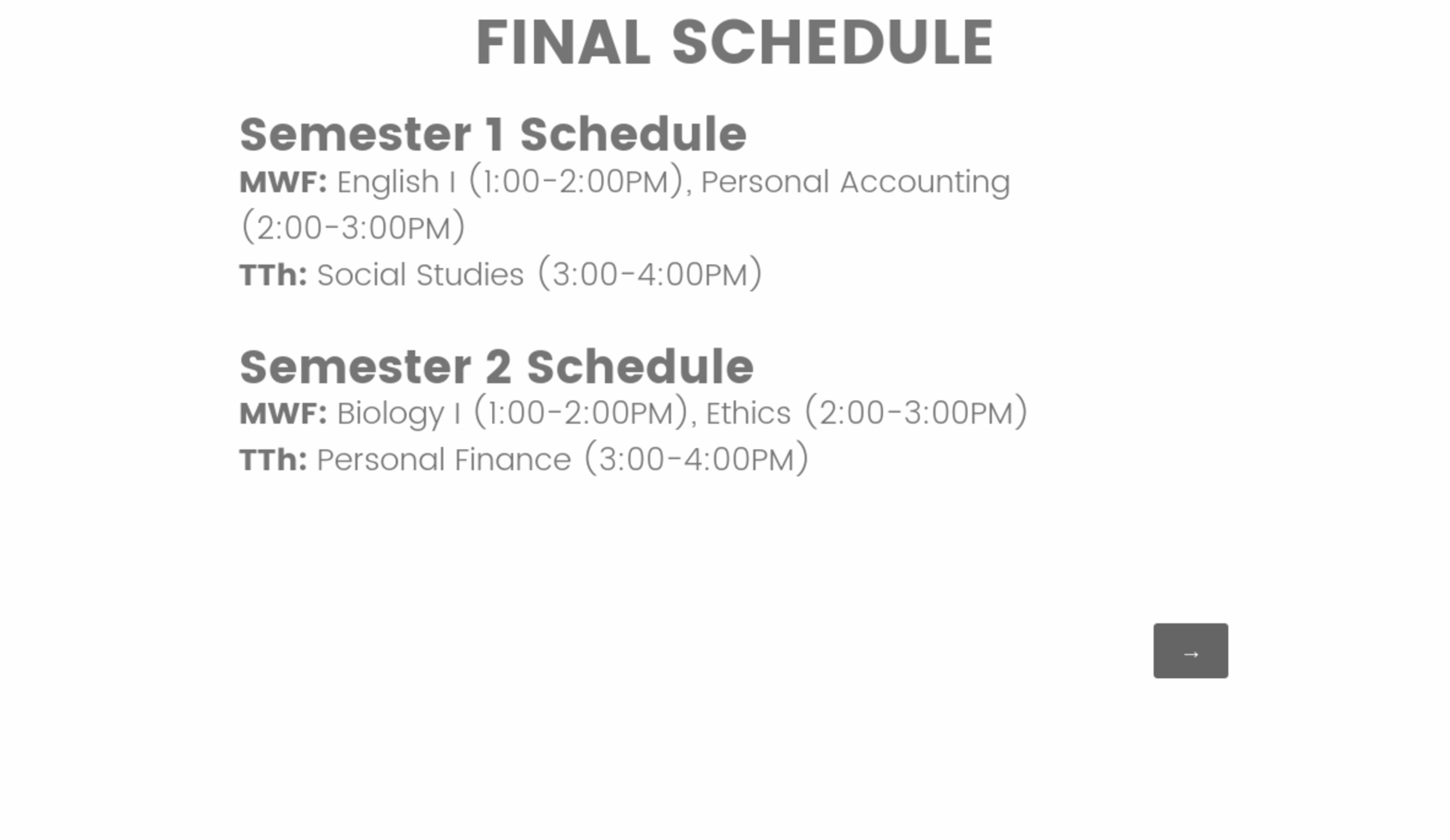}\hfill
  \makebox[0.47\linewidth]{} 
  
  \makebox[0.47\linewidth][c]{Step 11}\hfill\makebox[0.47\linewidth]{}

\caption{Screenshots from a mistake simulation shown in chronological order (Steps 5--11).}
\label{fig:appendix-sim-seq-part2}
\end{figure*}

\FloatBarrier

\section*{Acknowledgments}
\label{sec:ack}
Dr. Sreedharan's research is supported by NSF through grant 2303019. This material is based in part upon work supported by Other Transaction award 1AY2AX000062 from the U.S. Advanced Research Projects Agency for Health (ARPA-H) Platform Accelerating Rural Access to Distributed Integrated Medical Care (PARADIGM) program. The views and conclusions contained in this document are those of the authors and should not be interpreted as representing the official policies, either expressed or implied, of the U.S. Government.

\bibliography{references}

\end{document}